\documentclass[12pt]{article}

\usepackage{graphics}

\newcommand{\secn}[1]{Section~\ref{#1}}

\newcommand{\eq}[1]{Eq.~(\ref{#1})}
\newcommand{\fig}[1]{Fig.~\ref{#1}}
\newcommand{\nl}{\nonumber \\}

\def\beq{\begin{equation}}
\def\eeq{\end{equation}}
\def\beqa{\begin{eqnarray}}
\def\eeqa{\end{eqnarray}}

\newcommand{\sect}[1]{\setcounter{equation}{0}\section{#1}}
\renewcommand{\theequation}{\thesection.\arabic{equation}}
\newcommand{\EQ}{\begin{equation}}
\newcommand{\EN}{\end{equation}}
\newcommand{\bea}{\begin{eqnarray}}
\newcommand{\ena}{\end{eqnarray}}

\renewcommand{\a}{\alpha}
\renewcommand{\b}{\beta}
\renewcommand{\c}{\gamma}
\renewcommand{\d}{\delta}
\renewcommand{\l}{\lambda}

\renewcommand{\thefootnote}{\fnsymbol{footnote}}

\begin{document}
\begin{titlepage}
\rightline{DFTT 68/99}
\rightline{NEIP--99--022}
\rightline{\hfill December 1999}

\vskip 2.5cm

\centerline{\Large \bf Scalar field theory limits}
\centerline{\Large \bf of bosonic string amplitudes} 
 
\vskip 2cm

\centerline{\bf Alberto Frizzo and Lorenzo Magnea\footnote{e-mail: 
magnea@to.infn.it}}
\centerline{\sl Dipartimento di Fisica Teorica, Universit\`a di Torino}
\centerline{\sl and I.N.F.N., Sezione di Torino}
\centerline{\sl Via P.Giuria 1, I--10125 Torino, Italy}

\vskip .5cm
 
\centerline{\bf Rodolfo Russo}
\centerline{\sl Institut de Physique, Universit\'e de Neuch\^atel}
\centerline{\sl Rue A.-L. Breguet 1, CH--2000 Neuch\^atel, Switzerland}
 
\vskip 2cm
 
\begin{abstract}

We describe in detail the techniques needed to compute scattering
amplitudes for colored scalars from the infinite tension limit of
bosonic string theory, up to two loops. These techniques apply both to
cubic and quartic interactions, and to planar as well as non--planar
diagrams. The resulting field theories are naturally defined in the 
space--time dimension in which they are renormalizable. With a careful 
analysis of string moduli space in the Schottky representation we determine
the region of integration for the moduli, which plays a crucial 
role in the derivation of the correct combinatorial and color 
factors for all diagrams. 

\end{abstract}

\end{titlepage}

\newpage
\renewcommand{\thefootnote}{\arabic{footnote}}
\setcounter{footnote}{0}
\setcounter{page}{1}

\sect{Introduction}
\label{intro}
\vskip 0.5cm

It is well known that all perturbative states in string theory have a
squared mass proportional to the string tension $T$. Thus, in the low--energy 
regime (or zero--slope limit, $\a' = 1/(2\pi T) \to 0$), the heavy
string states become infinitely massive and decouple, while the light
states survive and their dynamics can be effectively described by an
ordinary field theory. It was understood, already in the old days of
dual models, that one can define this point--like limit in different ways,
so that it is possible to recover different field theories.
In the first application of this idea~\cite{scherk71}, the tree--level 
amplitudes of a scalar field theory with cubic interactions were derived 
from the corresponding amplitudes among scalar string states; it was then 
shown that, if massless spin--1 states are selected, in the low--energy limit
one can reproduce the tree diagrams of Yang--Mills theory~\cite{nesche72}, 
while, if closed string are considered, one obtains the amplitudes of 
Einstein's gravity~\cite{yo73,scheschw74}.

In more recent years further steps were taken, and string techniques were 
actually exploited as a {\em simplifying} tool that can substitute Feynman
diagrams for the explicit calculation of scattering amplitudes and other 
quantities of interest in field theory. For example, effective actions
and threshold effects of interest for string unification were computed 
in~\cite{mets,kap88,kap92}; string--inspired techniques were applied to 
the evaluation of one--loop QCD scattering 
amplitudes~\cite{beko91,beko92,be92,bediko93,bernrev} and renormalization 
constants~\cite{beko88,letter,big}; graviton scattering amplitudes were 
computed and their relation to gauge amplitudes explored~\cite{bedushi,bddpr};
progress was made towards the extension of the method to more than one 
loop~\cite{kaj,2loop,rolsat1,rolsat2,maru,berntwol}, to amplitudes with 
external fermions~\cite{rolpa}, and to off--shell amplitudes~\cite{nap}. 
String techniques also served to stimulate the development of new 
techniques in field theory, that preserve some of the nice features of the 
string formalism~\cite{bedu,strass,schu,schu2,schsat1,schsat2}.

At first sight, the use of extended object for constructing particle 
scattering amplitudes may appear as an unnecessary detour, since one is
interested only in the zero--slope limit. However, as is now well understood,
this procedure presents many useful features. For instance, string 
amplitudes are naturally written in a way that takes maximal advantage of 
gauge invariance, and the color decomposition is automatically performed.
At higher order in the perturbative expansion, further advantages become 
apparent: the loop momentum integrals are already performed, so that 
helicity methods can be readily employed, and the result for a set of 
Feynman diagrams of a given topology is presented directly as a 
Schwinger--parameter integral. Moreover, one does not find the 
large proliferation of diagrams characteristic of field theories, which 
makes it extremely difficult to perform high order calculations. 
In the case of closed strings one gets only one diagram at each order, 
while in the open string the number of diagrams remains small.
Finally, in the case of bosonic strings, the expressions of scattering
amplitudes and of the measure of integration on moduli space are
known explicitly for an arbitrary perturbative order; in the sewing
procedure~\cite{dfls88,d92}, they can be obtained from
tree level diagrams by identifying pairs of external legs with an 
appropriate propagator. 

On general grounds, a striking difference between field--theoretic and 
string--derived amplitudes is the degree of correlation in the calculation 
of different amplitudes in different field theories. In field theory,
different amplitudes are largely independent of each other, and thus in 
each computation one has to start from the same basic ingredients, the 
Feynman rules; also, results obtained in one theory can rarely be exploited
in computations for a different theory: in fact, introducing some 
modification in the defining Feynman rules, all subsequent results are 
affected. In the string approach, the situation is different: the basic
ingredients of all calculations always arise from quantities defined on 
the string world--sheet, and thus are not dependent on the specific 
definition of the field theory limit one may consider. In particular, the 
building blocks are the measure of integration over string moduli and 
the Laplacian Green function on the string world--sheet, essentially the 
two--point correlator of two--dimensional scalar fields.  This means that a 
large fraction of the calculation is done in a general framework, without 
specifying which states will be selected by the field theory limit. 
All these results can then be exploited for deriving different amplitudes in 
a given field theory, or even for calculations applying to different field
theories. For instance, the number of external particles does not play such 
a dramatic role in determining the complexity of the calculation. As we will 
show explicitly in the two--loop case, one can learn a lot of information 
about the shape of the relevant string world--sheets by studying
simple vacuum bubble diagrams. The results obtained in this way
will not be modified by the insertion of external legs.
Similarly, there are relations among calculations in different theories; 
in fact, one chooses a specific field theory only when 
one selects, in the string master formula, the contributions of a 
specific state. Technically this step simply amounts to a Taylor expansion 
of all functions appearing in the string amplitude, keeping the 
appropriate powers of the variables describing the string world--sheet. 
Clearly, the building blocks of the calculation are not modified by 
changing the string state one focuses on; also, the overall normalization 
and the integration region over string moduli are fixed once and for all.
In general, this unifying way of treating amplitudes of different
theories brings many simplifications; for instance, the tensor algebra 
associated with the propagation of spin--$1$ (or spin--$2$) particles is 
bypassed, and the computational complexity of these amplitudes is almost 
reduced to that of scalar amplitudes. 

Despite all the advantages just described, so far the technique has 
been fully applied only to massive scalars and to massless gauge bosons and 
gravitons, and has been completely successful only at one loop. There are 
several technical reasons for this limitation: the extension to fermions
requires in principle the use of superstring amplitudes, and the multiloop
technology in that case has not yet been completely developed; the 
extension to massive particles with spin is in principle possible, however 
one should realize that string theory is clearly ill--equipped to reproduce 
field theories with several mass scales, since all scales at the string 
level are proportional to the string tension $T$; finally, the extension 
to two and more loops has proved harder than expected, because it requires
a detailed understanding of multiloop string moduli space, whose
corners contributing to the field theory limit manage to reconstruct the
particle amplitudes in a highly non--trivial way. The present paper is a 
step towards the solution of the problems connected with the application 
of string techniques to multiloop diagrams. In particular, we study the 
two--loop open string moduli space, determine the correct integration 
region over moduli and punctures in the field theory limit, and show how
different corners of moduli space cooperate to reconstruct individual 
Feynman diagrams, with the correct symmetry and color factors.

In this paper we focus on the study of scalar interactions. This means
that both for external and for propagating states we will select the 
contributions coming from the open string tachyon, a Lorentz scalar 
taking values in the adjoint representation of the Chan--Paton group,
which we shall take to be $U(N)$. In \secn{mscal} we will introduce the 
technical tools needed for the computation of bosonic string amplitudes, 
and the Schottky parametrization of the string world--sheet. 
In \secn{tree} we will show how one can define two different point--particle 
limits, by matching in different ways the field coupling constant 
$g$, which is kept fixed when $\a'\!\to\! 0$, to the unique string 
coupling $g_S$. The two matchings lead to a cubic scalar interaction in
$d = 6 - 2 \epsilon$, and to a quartic scalar interaction in 
$d = 4 - 2 \epsilon$, respectively. Each field theory naturally arises in 
the space--time dimension in which it is renormalizable, essentially 
because the string scale in the intermediate stages plays the role
of a renormalization scale, and disappears from the  matching conditions
when the field coupling becomes dimensionless. As a first check, some
tree--level amplitudes are derived from the string master formula.
In \secn{oneloop} we turn to one--loop diagrams, which are fairly 
straightforward to handle, and thus serve as a useful preliminary to 
two--loop calculations. We give explicit expressions for multi--leg 
one--particle--irreducible diagrams both for cubic and quartic 
interactions, and we show how non--planar color structures (subleading in 
the large--$N$ limit) are correctly reproduced in the string framework.
Finally, in \secn{twoloop}, we turn to the more challenging problem of 
two--loop diagrams. Different calculations are presented, up to the four--point 
amplitude, both for cubic and quartic interactions. We present a detailed 
analysis of the two--loop moduli space in the field theory limit, which 
leads us to recover the correct results {\em including} the normalization 
factors\footnote{A different method to identify the regions of moduli space
corresponding to two--loop quartic interactions, and to compute the 
corresponding amplitudes, has recently been introduced in Ref.~(\cite{MP}).}. 
Indeed, it is clear that for these more complicated diagrams 
the color decomposition and the combinatoric coefficients are obtained 
more easily from the string approach than from usual field theory 
techniques. In \secn{concl} we present our conclusions, and our current 
assessment of the status of our method.

\sect{Multiloop scalar amplitudes in string theory}
\label{mscal}
\vskip 0.5cm

\begin{figure}[ht] 
\begin{center} 
{\scalebox{0.55}[0.75]{\includegraphics{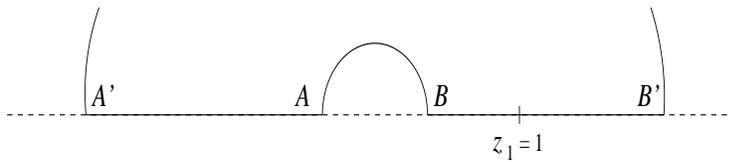}}}
\end{center} \caption{The annulus in the Schottky representation.}
\label{sch1}
\end{figure}

As we have already anticipated, in bosonic string theory it
is possible to write in a compact form a generic loop amplitude 
among string states, with an arbitrary number of loops and external legs. 
This formula can be immediately specialized to the case where all 
external states are scalars. The full, normalized, $h$--loop scattering 
amplitude~\cite{2loop} of $M$ tachyons with momenta $p_1,\ldots,p_M$,
can be written as
\bea
A^{(h)}_M (p_1,\ldots,p_M) & = & 
\prod_{r = 1}^{h + 1} \Big({\rm Tr}\,(\lambda^{a_1^r} 
\cdots \lambda^{a_{N_r}^r}) \Big)~
\frac{2^M ~g_S^{M + 2 h - 2}}{(2\pi)^{d h}} 
\left( 2 \a' \right)^{(M d - 2 M - 2 d)/4}
\nl & \times &
\int [dm]^M_h \,
\prod_{i<j} \left[\frac{ \exp \left({\cal G}^{(h)}_{r_i,r_j} (z_i,z_j) 
\right)}{\sqrt{V'_i(0) \, V'_j(0)} } \right]^{2 \a' p_i\cdot p_j}~~.
\label{hmastac}
\ena
Here $g_S$ is the dimensionless string coupling constant, and
the product of traces is the appropriate $U(N)$ Chan--Paton factor for a 
generic $h$--loop diagram with $h + 1$ boundaries labelled by the index $r$.
In the planar case it becomes simply $N^h \, {\rm Tr}(\lambda^{a_1} 
\cdots \lambda^{a_M})$. ${\cal G}^{(h)}_{r_i,r_j}$ is the correlator of 
two world--sheet bosons located at $z_i$ on the boundary labelled $r_i$, and
at $z_j$ on the boundary $r_j$, while $[dm]^M_h$ is the measure of integration
over moduli space for an open Riemann surface with $h$ loops and $M$ 
punctures. Notice that since we consider $U(N)$ as a gauge group, 
we have to take into account only string amplitudes with oriented 
world--sheets; thus non--planar diagrams arise only when loops are
formed by sewing together two non--consecutive punctures.
Here we will not be interested in the exact expression of the geometric
objects appearing in \eq{hmastac}, which can be found in~\cite{scho}; rather,
we will focus on their general features, in order to emphasize the properties 
which play a crucial role in the field theory limit. For a more complete 
presentation of the mathematical tools we will briefly describe here, we 
refer to Ref.~\cite{scho}.
As is well known, the Schottky parametrization is particularly suited 
for the study of the field theory limit, so we will work always within this 
framework. In Fig.~\ref{sch1}, in \secn{oneloop}, 
and Fig.~\ref{2lf}, in \secn{twoloop}, we present 
the one-- and the two--loop string world--sheets in the Schottky 
representation. It is easy to see how the idea of adding loops is 
implemented in this formalism. One starts from the upper
half complex plane (equivalent to the disk, representing the tree--level 
scattering amplitude), and adds two circles with the same 
radius and with centers on the real axis, which must then be identified via a
projective transformation. Each loop is thus characterized by three real 
parameters: the positions of the two centers on the real axis, and the 
radius of the circles, which fix respectively the position and the width of 
the holes added to the surface. The positions of the various circles are 
related to the fixed points of the projective transformations under which the 
pairs of circles are identified, usually denoted by $\xi_\mu$ and $\eta_\mu$,  
($\mu = 1, \ldots, h$), while the width of the holes is determined by the 
third parameter characterizing the projective transformation, the multiplier
$k_\mu$. It is possible to use the projective invariance of string theory
to fix the location of up to three of the punctures $z_i$, or of the 
fixed points, but one cannot fix the multipliers, which in fact drive the 
field theory limit. In general, it is convenient to fix, say, two $\xi$'s and 
one $\eta$, except in the one--loop case, where one has only two fixed 
points; in this case one also specifies the position of one of the punctures.
In the field theory limit, the surface must degenerate into a graph, and all
massive string modes must decouple. One can show in general~\cite{kaj} that
the relevant region of string moduli space is the region $k_\mu \to 0$, and 
furthermore the Taylor expansion of the integrand of the string amplitude
in powers of the multipliers corresponds to a sum over the mass levels of 
the states circulating in the loops. Thus, for the field theory limit of 
scalar amplitudes, we can always ignore all higher powers in the multipliers.
In this limit, the integration measure in \eq{hmastac} reads
\beq
[dm]^M_h =  \Delta(\rho_a, \rho_b, \rho_c) ~
\prod_{\mu=1}^{h} \left[ \frac{dk_\mu \,d \xi_\mu \,d \eta_\mu}{k_\mu^2
\,(\xi_\mu - \eta_\mu)^2}\right]
\left[\det \left( - i \tau_{\mu \nu} \right) \right]^{-d/2}
\left( \prod_{i=1}^M \frac{dz_i}{ V_{i} ' (0)}\right)~, \nonumber
\label{hmeasure}
\eeq
where the factor $\Delta(\rho_a, \rho_b, \rho_c)$ is the Faddeev--Popov
determinant associated with the fixing of the overall projective
invariance, and $\tau_{\mu \nu}$ is the period matrix of the surface, whose 
explicit expression at one and two loops will be given in Eqs.~(\ref{multi3}) 
and (\ref{permat}) below. Note that all
the dependence on the external states is concentrated
in the last term. The factors ${V_{i}'(0)}$ originate from the need 
to introduce local coordinates on the surface, $V_i(z)$, around each 
puncture, in order to perform the sewing procedure. 
Before discussing their role, let us
introduce explicit expressions for the Green functions we will need. Also
in this case, we report here just the leading term in the Taylor expansion 
in the multipliers (see~\cite{scho} for the complete string expressions).
At one loop, if the two punctures are on the same boundary, 
one finds~\cite{big}
\beq
{\cal G}^{(1)} \left(z_i, z_j\right) = \log \left| z_i -z_j\right| 
+ \frac{1}{2 \log k} \log^2 \frac{z_i}{z_j} ~~,
\label{Gr1}
\eeq
otherwise one must use the ``non-planar'' Green function
\beq
{\cal G}_{NP}^{(1)} \left(z_i, z_j\right) = \log \left| z_i +z_j\right| 
+ \frac{1}{2 \log k} \log^2 \frac{|z_i|}{|z_j|} ~~.
\label{Gr1np}
\eeq
Notice that in Eqs.~(\ref{Gr1}) and (\ref{Gr1np}) we have already
chosen the projective gauge $\eta = 0$ and $\xi \to \infty$, so that only
the multiplier $k$ appears explicitly. 
At two loops the bosonic Green function becomes a little more
complicated because one has to deal with a non--trivial 
dependence on the moduli of the two holes. In the planar case
\bea
{\cal{G}}^{(2)} (z_1 , z_2) & = & \log | z_1 - z_2 | + \frac{1}{2}
\left[ \log k_1  \log k_2 -  \log^2 S \right]^{-1} \label{Green} \\
& \times & \left[ \log^2 T \log k_2 + \log^2 U \log k_1  - 2 \log T \log U
\log S \right]~~.   \nonumber
\ena
The Green function now depends on two different multipliers and
on four fixed points through the anharmonic ratios
\bea
S & = & \frac{(\eta_1 - \eta_2 ) (\xi_1 - \xi_2 )}{(\xi_1 - \eta_2 )
(\eta_1 - \xi_2 )}~~, \nonumber \\
T & = & \frac{(z_2 - \eta_1 ) (z_1 - \xi_1 )}{(z_2 - \xi_1 )
(z_1 - \eta_1 )}~~,  \label{STU}  \\
U & = &\frac{(z_2 - \eta_2 ) (z_1 - \xi_2 )}{(z_1 - \eta_2 )
(z_2 - \xi_2 )}~~. \nonumber
\ena
As was already noticed in~\cite{big}, at one loop, these
Green functions do not have the expected periodicity properties.
This is not really surprising, since it is known that the factor 
$\exp[{{\cal G}(z_i,z_j)}]$ appearing in the master equation (\ref{hmastac})
has conformal weight ($- 1/2$, $- 1/2$) in the two variables 
($z_i$,$z_j$) and not zero. Thus to give a global definition to 
\eq{hmastac}, one should multiply it by a function of conformal 
weight $(1/2,1/2)$. This suggests that one can recover a well behaved 
geometric object if the local coordinates ${ V_{i}'(0) }$ compensate 
for this problem by having conformal weight $-1$. A natural choice is to 
define the ${ V_{i}'(0)}$ by using the inverse of the abelian differentials, 
which are the only globally defined objects having conformal weight one. By
following this idea one recovers at one loop the choice made in
\cite{big} where ${ V_{i}'(0)} = \omega^{-1}(z_i) = z_i$, since, in this
case, there is a unique abelian differential, $\omega(z) = 1/z$.  
At two loops, one is lead to identify the inverse of ${ V_{i}'(0)}$ with a 
linear combination of the two differentials $\omega_1(z_i)$ and 
$\omega_2(z_i)$; to fix the normalization, we will follow 
Ref.~\cite{rolsat2}, and require that this linear combination 
be normalized to one when one integrates it around the field theory 
propagator on which the leg is inserted. This is sufficient to fix the 
local coordinates for the purpose of the scalar field theory limit.

The string expressions reported here contain just the leading order
in the multipliers; moreover, as we anticipated in the Introduction, in
order to derive the field theory limit of \eq{hmastac}, one has to
expand the integrand also in powers of all the other variables. The logarithmic
terms, however, are non--analytic and must have a special status. In
fact, it turns out that they measure the length of field theory 
propagators in units of $\a'$, and so are directly related to the
Schwinger proper times of the limiting field theory. This means that, 
technically, the zero--slope limit has to be taken after introducing 
the field theory variables that have to be kept fixed: the string coupling 
constant has to be translated to the appropriate field theory coupling, 
while the logarithmic terms in the integrand must be interpreted in terms of 
proper times, in general as $\ln{(x)} \propto t/ \a' $. The exact form of
this change of integration variables depends on the particular corner of
moduli space considered, as we will see in detail in the following sections.

Although we gave explicit expressions for all functions entering in
\eq{hmastac}, our master formula is still a formal expression, since we 
have not yet specified the exact region of integration of the various 
parameters. We do not attempt to solve this problem in its most general form; 
however, in the following sections, we will determine the correct
region of integration, at least in the field theory limit, 
for the one-- and two--loop diagrams. Here we just anticipate the basic idea. 
One starts by considering the vacuum amplitudes, where it is possible
to focus only on the world--sheet shape, without having to consider external
punctures. In this way one determines the region of integrations over
the fixed points, by requiring that the surface never become singular;
on the other hand, the integration over the multipliers is fixed by
symmetry arguments and is chosen in order to avoid double counting of
equivalent  configurations. When external legs are added, this setup
is not essentially modified. From this analysis of the string
world--sheet one can draw a clear representation of the various
boundaries of the surface, and the  punctures can take all possible
values on these boundaries.

\sect{Tree--level matching conditions}
\label{tree}
\vskip 0.5cm

We begin our analysis of the field theory limit by establishing the
relationships between the string coupling and the couplings of the
cubic and quartic theories we want to reproduce. This is done, as in
effective field theory, by computing the simplest amplitudes both with
strings and fields, and matching the results. The string amplitude is,
of course, uniquely defined: different matchings correspond to
different ways of taking the infinite tension limit, and they lead to
different field theories in different dimensions. Having established 
the connection between the couplings, we go on to describe the computation 
of simple tree diagrams, with up to six external legs.

The on--shell, tree--level, color--ordered, $M$--point scalar 
amplitude in bosonic string theory is readily derived from
\eq{hmastac}, by choosing as Green function simply ${\cal
G}(z_i,z_j) = \ln{|z_i-z_j|}$. The result is the correctly 
normalized Koba--Nielsen amplitude,
\beqa
A_M^{(0)} \left(p_1, \ldots , p_M \right) & = & {\rm Tr} \left(
\l^{a_1} \ldots \l^{a_M} \right) ~2^M ~g_S^{M - 2} 
(2 \a')^{(M d - 2 M - 2 d)/4} \nonumber \\ & \times & ~\int 
\prod_{i = 1}^M d z_i ~\Delta (z_a, z_ b, z_c) ~\prod_{i < j}
(z_i - z_j)^{2 \a' p_i \cdot p_j}~~.
\label{treemast}
\eeqa
Here the punctures $z_i$ are ordered on a circle, as in the trace, and
$\Delta(z_a, z_b, z_c)$ is the Faddeev--Popov determinant arising
from the fixing of projective invariance,
\beq
\Delta(z_a, z_b, z_c) = \d(z_a - z_a^{(0)}) 
\d(z_b - z_b^{(0)}) \d(z_c - z_c^{(0)}) (z_a - z_b) 
(z_a - z_c) (z_b - z_c)~~,
\label{fapop}
\eeq
where $z_{a,b,c}$ are three arbitrarily chosen punctures whose
location is fixed. In the present section we will always choose
$z_1 \to  \infty$, $z_2 = 1$ and $z_M = 0$, so that all remaining
integrals range between $0$ and $1$.

With these choices, the $3$--point amplitude is simply
\beq
A_3^{(0)} \left(p_1, p_2, p_3 \right) = {\rm Tr} \left(\l^{a_1} 
\l^{a_2} \l^{a_3} \right) ~8 g_S ~(2 \a')^{(d - 6)/4}~~,
\label{3string}
\eeq
whereas the $4$--point amplitude (contributing to the Veneziano
formula) is given by
\beqa
A_4^{(0)} \left(p_1, p_2, p_3, p_4 \right) & = & {\rm Tr} \left(\l^{a_1} 
\l^{a_2} \l^{a_3} \l^{a_4} \right) ~16 g_S^2 ~(2 
\a')^{(d - 4)/2} \nonumber \\ 
& \times & \int_0^1 d z z^{2 \a' p_3 \cdot p_4}
(1 - z)^{2 \a' p_2 \cdot p_3}~~.
\label{4string}
\eeqa
Since we wish to consider a slightly more complicated example, we also
give the expression for the $6$--point function,
\beqa
A_6^{(0)} \left(p_1, \ldots , p_6 \right) & = & {\rm Tr} \left(\l^{a_1} 
\ldots \l^{a_6} \right) ~64 g_S^4 ~(2 \a')^{(d - 3)} 
\int_0^1 d z_3 \int_0^{z_3} d z_4 \int_0^{z_4} d z_5 
\nonumber \\ \times & & \! \! \! \! \! \! \! \! \! \! \! \!
\left[ (1 - z_3)^{2 \a' p_2 \cdot p_3} (1 - z_4)^{2 \a' p_2 
\cdot p_4} (1 - z_5)^{2 \a' p_2 \cdot p_5} (z_3 - z_4)^{2 \a' 
p_3 \cdot p_4} \right. \nonumber \\ 
\times & & \! \! \! \! \! \! \! \! \! \! \! \!  \left.
(z_3 - z_5)^{2 \a' p_3 \cdot p_5} (z_4 - z_5)^{2 \a' p_4 
\cdot p_5} z_3^{2 \a' p_3 \cdot p_6} z_4^{2 \a' p_4 \cdot p_6}
z_5^{2 \a' p_5 \cdot p_6} \right]~~.
\label{6string}
\eeqa
Turning to field theory, let us first consider a scalar $U(N)$ theory
defined by the lagrangian\footnote{Note that here we use a convention slightly
different from the one employed in Ref.~\cite{2loop}; in particular we
choose a coupling constant $g_3$ which is four times bigger than the
one of \cite{2loop}, in order to reduce the number of factors of two in
the amplitudes.}
\beq
{\cal L}_3 = {\rm Tr} \left[ \partial_\mu \Phi \partial^\mu \Phi - m^2
\Phi^2 + \frac{2}{3} g_3 \Phi^3 \right]~~,
\label{lag3}
\eeq
where $\Phi = \phi_\a \l^\a$ is a scalar field taking
values in the adjoint representation of $U(N)$, and our generators are
normalized by ${\rm Tr} (\l_\a \l_\b) = \d_{\a \b}/2$. 
Feynman rules for scalar $U(N)$ theories and several useful formulas
for color structures are collected in the Appendix. The color--ordered
$3$--point amplitude (defined as $- {\rm i}$ times the relevant
Feynman diagram) in this theory is simply
\beq
A_3^{(0)} \left(p_1, p_2, p_3 \right) = 2 g_3 ~{\rm Tr} \left(\l^{a_1} 
\l^{a_2} \l^{a_3} \right)~~.
\label{3field}
\eeq
Comparison with \eq{3string} yields the matching condition
\beq
g_3 = 4 g_S ~(2 \a')^{(d - 6)/4}~~,
\label{match3}
\eeq
already derived in~\cite{2loop}. Note that the coupling is
dimensionless in $d = 6$, as it must.
We can now use \eq{match3} to compute higher order scattering
amplitudes in the theory defined by \eq{lag3}, using the string master
formula. As a first simple example, consider the $4$--point amplitude,
which becomes
\beqa
A_4^{(0)} \left(p_1, \ldots , p_4 \right) & = & {\rm Tr} \left(\l^{a_1} 
\ldots \l^{a_4} \right) ~2 g_3^2 ~\a' \nonumber \\ 
& \times & \int_0^1 d z z^{2 \a' p_3 \cdot p_4}
(1 - z)^{2 \a' p_2 \cdot p_3}~~.
\label{4first}
\eeqa
To get a sensible zero slope limit we must extract from the integral a
factor $(\a')^{-1}$. This can be done by focusing on the
potentially singular regions $z \to 0$ and $z \to 1$, corresponding to
$s$-- and $t$--channel exchange respectively (the $u$--channel
diagram cannot contribute to this color structure). According to the
general discussion of \secn{mscal}, and considering the region $z \to
0$, we do this by setting $z = \exp (- t_z/\a')$, with $t_z$ finite
and $\a' \to 0$. Neglecting $O(\a')$ corrections, the change
of variables yields
\beqa
A_4^{(0)} \left(p_1, \ldots , p_4 \right) & = & {\rm Tr} \left(\l^{a_1} 
\ldots \l^{a_4} \right) ~2 g_3^2 ~\int_0^\infty d t_z \exp 
\left[- \frac{t_z}{\a'} \left(1 + 2 \a' p_3 \cdot p_4 \right)
\right] \nonumber \\
& = & 2 g_3^2 ~{\rm Tr} \left(\l^{a_1} \ldots \l^{a_4} 
\right) ~\frac{1}{(p_3 + p_4)^2 + m^2}~~,
\label{4sec}
\eeqa
where we made use of the mass--shell condition $m^2 = -
1/\a'$. Taking into account the similar contribution arising from
the region $z \to 1$, we get the complete answer for this color
ordering,
\beq
A_4^{(0)} \left(p_1, \ldots , p_4 \right) = 2 g_3^2 ~{\rm Tr} 
\left(\l^{a_1} \ldots \l^{a_4} \right) 
\left[ \frac{1}{(p_3 + p_4)^2 + m^2} + \frac{1}{(p_2 + p_3)^2 + m^2} 
\right]~~,
\label{4fin}
\eeq
which exactly matches the correct result in field theory, provided the
string metric is used.

\begin{figure}[ht] 
\begin{center} 
{\scalebox{0.5}{\includegraphics{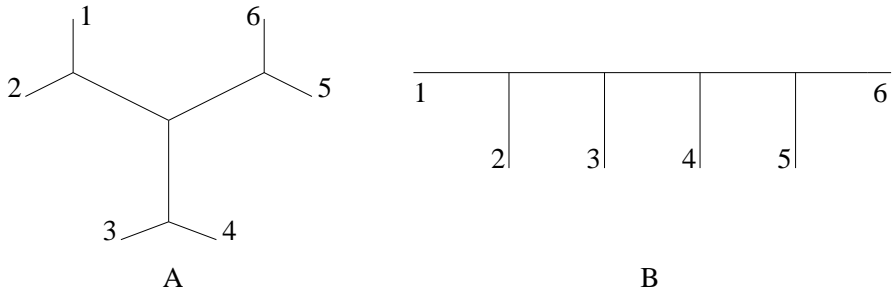}}}
\end{center} \caption{Two tree--level six--point diagrams with cubic 
vertices.}
\label{btree}
\end{figure}

It is clear that the different diagrams contributing to a given color
ordering arise from different corners of string moduli space, and
they are easily identified by the pole structure of their
propagators. To illustrate this in a slightly less trivial
configuration, let us consider the $6$--point amplitude in
\eq{6string}, and let us attempt to separate the contributions to two
different given diagrams, say those portrayed in Fig.~\ref{btree}.
Using the matching condition, \eq{match3}, one easily sees that the 
string amplitude is proportional to $(\a')^3$, so that we need to take 
a limit in all three integration variables, which corresponds to the 
extraction of three propagator poles. If we want to specify the region in 
moduli space relevant for diagram $A$, we must qualitatively have 
$z_5$ very close to $z_6 = 0$, and $z_3$ very close to $z_4$. Moreover the 
three pairs of variables $(z_1, z_2)$, $(z_3, z_4)$ and $(z_5, z_6)$ 
have to be kept widely separated from each other.
The desired change of variables is
\beqa
z_3 & = & {\rm e}^{- t_3/\a'} \nl
z_3 - z_4 & = & {\rm e}^{- t_4/\a'} \label{var6}\\
z_5 & = & {\rm e}^{- t_5/\a' \nonumber}~~.
\eeqa
In terms of these ``proper times'', and neglecting 
terms suppressed by powers of $\a'$, the amplitude reads
\beqa
& & A_6^{(0)} \left(p_1, \ldots , p_6 \right) ~=~ {\rm Tr} \left(\l^{a_1} 
\ldots \l^{a_6} \right) ~2 g_3^4 ~\int_0^\infty d t_3
\int_{t_3}^\infty d t_4 \int_{t_3}^\infty d t_5 
\nl & & ~\times \exp \left[- \frac{t_4}{\a'} (1 + 2 \a' p_3
\cdot p_4) \right] ~\exp \left[- \frac{t_5}{\a'} (1 + 2 \a'
p_5 \cdot p_6) \right] \nl
& & ~\times \exp \left[- \frac{t_3}{\a'} (1 + 2 \a'
p_3 \cdot p_5 + 2 \a' p_3 \cdot p_6 + 2 \a' p_4 \cdot p_5 + 
2 \a' p_4 \cdot p_6 ) \right]~.
\label{6out}
\eeqa
One easily sees that the contribution of this region to the 6-point 
amplitude is
\beq
A_6^{(0)} ~= ~{\rm Tr} \left(\l^{a_1} \ldots \l^{a_6} \right) ~2
g_3^4 ~\frac{1}{(p_1 + p_2)^2 + m^2} ~\frac{1}{(p_3 + p_4)^2 + m^2}~
\frac{1}{(p_5 + p_6)^2 + m^2}~~,
\label{sixfin}
\eeq
precisely the desired result. 

For the other diagram of Fig.~\ref{btree}, one has to consider a different 
change of variables; in particular, since now we do not want to group
external particles in pairs, all $z_i$'s must be taken widely separated. Thus 
the corresponding ``ordered proper times'' are simply defined as $t_i =
-\a' \ln{z_i}$. Following the same steps just described, it is the easy
to check that the expected result for diagram $B$ is obtained.
We note in passing that finding the numerical coefficient of a given color 
ordering for a given Feynman diagram may be a rather cumbersome task with the
conventional Feynman rules, whereas it is immediate here.

Let us now turn our attention to quartic interactions. We want to
reconstruct the amplitudes derived from the lagrangian
\beq
{\cal L}_4 = {\rm Tr} \left[ \partial_\mu \Phi \partial^\mu \Phi - m^2
\Phi^2 + g_4 \Phi^4 \right]~~,
\label{lag4}
\eeq
which, in particular, yields the color--ordered vertex
\beq
A_4^{(0)} \left(p_1, p_2, p_3, p_4\right) = 4 g_4 ~{\rm Tr} 
\left(\l^{a_1} \l^{a_2} \l^{a_3} \l^{a_4} \right)~~.
\label{4field}
\eeq
The starting point is now \eq{4string}, where however in this case we
do not need to generate any extra powers of $\a'$, as was done to
go from \eq{4first} to \eq{4sec}. Here the overall dimensionality is
correct, so all we need to do is take the $\a' \to 0$ limit and integrate
over $z$ without introducing any weight in special corners of moduli
space. In other words, the quartic vertex arises from integration over
finite regions of moduli space, and not from its boundaries (regions
of infinitesimal size in the field theory limit). As $\a' \to 0$,
the integrand of \eq{4string} simply becomes $1$, so we get
\beq
A_4^{(0)} \left(p_1, p_2, p_3, p_4 \right) = 16 g_S^2 ~(2 
\a')^{(d - 4)/2} ~{\rm Tr} \left(\l^{a_1} 
\l^{a_2} \l^{a_3} \l^{a_4} \right)~~.
\label{4lim}
\eeq
Comparison with \eq{4field} yields the matching condition
\beq
g_4 = 4 g_S^2 ~(2 \a')^{(d - 4)/2}~~.
\label{match4}
\eeq
The same matching condition might have been obtained in a different
way, by first considering the $\Phi^3$ diagrams contributing to
$A_4^{(0)}$, and then explicitly deleting the internal propagators by
setting their proper times to $0$, inserting in turn a
$\d(t_z/\a')$ and a $\d((t - t_z)/\a')$. These $\d$
functions should however be regularized, since they are located at the
boundaries of the integration region. In such circumstances it is
natural to weigh each $\d$ function with a factor $1/2$, and this
reproduces the matching in \eq{match4}. This second method is closer in 
spirit to the techniques of~\cite{MP}.

\begin{figure}[ht] 
\begin{center} 
{\scalebox{0.5}{\includegraphics{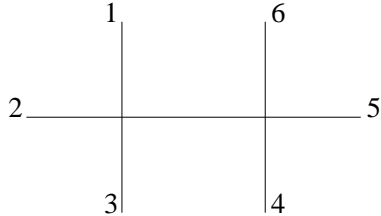}}}
\end{center} \caption{A sample tree--level six--point diagram with quartic 
vertices.}
\label{Gate}
\end{figure}

To show that this procedure can be generalized to higher order
amplitudes, let us consider also for $\Phi^4$ a particular diagram
contributing to the $6$--point amplitude. Note that there are $105$
diagrams contributing to the $\Phi^3$ $6$--point amplitude, but only
$10$ with quartic vertices. We consider, as an example, the diagram
depicted in Fig.~\ref{Gate}.
Using the matching condition \eq{match4} in
the string amplitude, \eq{6string}, one sees that the amplitude has an
overall factor of $\a'$. Thus we need to take precisely one
singular limit in one of the integration variables, corresponding to
the extraction of a single propagator pole. For the particular diagram
at hand, qualitatively, we would like to keep $z_4$ and $z_5$ close to
$z_6 = 0$, while leaving $z_3$ close to $1$. This can be achieved by
the change of variables
\beqa
z_3 & = & x \nl
z_4 & = & {\rm e}^{- t_4/\a'} \\ \label{var64}
z_5 & = & y ~{\rm e}^{- t_4/\a'}~~, \nonumber
\eeqa
with no proper times associated with $x$ and $y$. Neglecting as usual
terms suppressed by powers of $\a'$, we get
\beqa
A_6^{(0)} \left(p_1, \ldots , p_6 \right) & = & {\rm Tr} \left(\l^{a_1} 
\ldots \l^{a_6} \right) ~8 g_4^2 ~\int_0^1 d x ~\int_0^1 d y
~\int_0^\infty d t_4 \nl
& & \exp \left[- \frac{t_4}{\a'} (2 + 2 \a'
p_4 \cdot p_5 + 2 \a' p_4 \cdot p_6 + 2 \a' p_5 \cdot p_6 
\right] \nl
& = & {\rm Tr} \left(\l^{a_1} \ldots \l^{a_6} \right) ~8
g_4^2 ~\frac{1}{(p_4 + p_5 + p_6)^2 + m^2}~~. \label{64out}
\eeqa
Once again, the computation of the color factor of the corresponding 
Feynman diagram yields the same result.

\sect{One--loop diagrams}
\label{oneloop}

The multi--tachyon planar one--loop amplitude derived from \eq{hmastac} can 
be written as
\beqa
A^{(1)}_M (p_1,\ldots,p_M) & = & N \, {\rm Tr}(\lambda^{a_1} \cdots 
\lambda^{a_M}) ~\frac{1}{(4\pi)^{d/2}} ~2^M ~g_S^M ~
(2 \a')^{(M d - 2 M - 2 d)/4} \nl
& \times & \int_{0}^{1}  \frac{dk}{k^2} \,
\left( - \frac{\log k}{2} \right)^{-d/2} \int_{k}^{1} d z_{2} \int_{k}^{z_2}
d z_{3} \ldots \int_{k}^{z_{M - 1}} d z_M \nl
& \times & \prod_{i=1}^{M} \frac{1}{V_{i}'(0)} ~\prod_{i<j} 
\left[{{\exp \left({\cal G}^{(1)}(z_i,z_j) \right)}
\over{\sqrt{ V'_i(0)\, V'_j(0)}}}\right]^{2 \a' p_i \cdot p_j}~~,
\label{onemast}
\eeqa
where we neglected $O(k)$ terms in the measure of integration that 
will not contribute to the field theory limit, and we have introduced the 
local coordinates $V_i(z)$, according to the general discussion of 
\secn{mscal}. Projective invariance has been used to choose the fixed points
of the single Schottky generator as $\eta = 0$ and $\xi \to \infty$, and to 
fix $z_1 = 1$. In this configuration the world--sheet of the string (an
annulus) can be represented as in Fig.~\ref{sch1}. 
\vspace{0.2cm}
For a planar configuration, all punctures are constrained to lie 
on the same boundary of the string world sheet, thus, 
having fixed $z_1 = 1$, all other $z_i$ should 
be integrated over the interval $ B = \sqrt{k} < z_i < B' = 1/\sqrt{k}$, 
with the restriction on the ordering implied by the color trace. 
This would complicate the calculation of the field theory limit, 
since there would be contributions both from $z_i \to \sqrt{k} \to 0$ 
and from $z_i \to 1/\sqrt{k} \to \infty$.
It is possible to bypass this practical difficulty by making use of the fact 
that the string integrand is modular invariant, which in particular implies 
that the interval $[1,1/\sqrt{k}]$ can be mapped onto the interval 
$[k,\sqrt{k}]$. In fact, defining the effective one--loop Green function by
\beq
G^{(1)} (z_i, z_j; k) = {\cal G}^{(1)}(z_i,z_j) - \frac{1}{2} \log V_i'(0)
- \frac{1}{2} \log V_j'(0)~~,
\label{effgr1}
\eeq
 and choosing $V_i'(0) = z_i$, according to our general discussion, it is easy 
to check that the effective Green function at the string level is a function 
only of the ratio $\rho_{i j} = z_i/z_j$, and satisfies
\beqa
G^{(1)} \left(\rho_{j i}, k \right) & = & G^{(1)} \left(\rho_{i j}, k 
\right) \nl 
G^{(1)} \left( k \rho_{j i}, k \right) & = & G^{(1)} \left(\rho_{i j}, k 
\right)~~.
\label{modg1}
\eeqa
Using these properties, one can map all configurations with a subset of 
punctures in the interval $[1,1/\sqrt{k}]$ to configurations in which those
punctures have been moved to the interval $[k,\sqrt{k}]$, preserving the
ordering on the circle. This procedure yields the integration region in
\eq{onemast}. In the field theory ($k \to 0$) limit the effective Green 
function has the form
\beq
G^{(1)} \left(z_i, z_j; k \right) = \log \left(\sqrt{\frac{z_i}{z_j}} -
\sqrt{\frac{z_j}{z_i}} \right) + \frac{1}{2 \log k} \log^2 \frac{z_i}{z_j}~~,
\label{gr1}
\eeq
for $z_i > z_j$.

The generalization of \eq{onemast} to the case of non--planar diagrams is
known \cite{GSW}, and easily understood. A non planar diagram has punctures 
on both boundaries, so the factor of $N = {\rm Tr} {\bf 1}$ is replaced by 
the trace of the Chan--Paton factors corresponding to the punctures on the
second boundary (the interval $[-1/\sqrt{k}, - \sqrt{k}]$ in 
Fig.~\ref{sch1}). The region of integration of the corresponding $z_i$ is
an ordered region on the negative real axis. The string Green function 
involving two punctures on the negative real axis is precisely the same as the
one discussed above, since it is a function only of the ratio $\rho_{i j}$, 
which remains positive when both $z$'s change sign. The only subtlety involves
the Green function connecting punctures on different boundaries. In this case
the terms in the Green function arising from loop momentum integration behave 
differently, and one should choose $V_i'(0) = |z_i|$, so that the Green 
function remains real, and does not have any singularity when 
$|z_i| \to |z_j|$. In the field theory limit one finds simply
\beq
G^{(1)}_{NP} \left(z_i, z_j; k \right) = \log \left(
\sqrt{\frac{|z_i|}{|z_j|}} +
\sqrt{\frac{|z_j|}{|z_i|}} \right) + \frac{1}{2 \log k} 
\log^2 \frac{|z_i|}{|z_j|}~~,
\label{gr1np}
\eeq
where the two $z$'s have opposite signs.

\subsection{One--loop cubic interactions}
\label{olc}

\begin{figure}[ht] 
\begin{center} 
{\scalebox{0.5}{\includegraphics{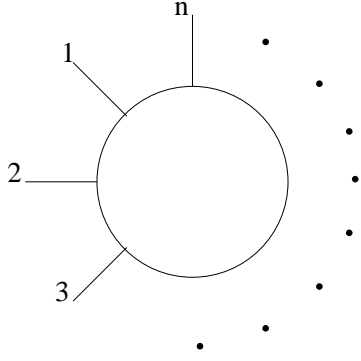}}}
\end{center} \caption{Multi--leg one--loop one--particle--irreducible diagram
in a $\Phi^3$ theory.}
\label{Sun}
\end{figure}

Armed with the appropriate string technology, let us examine how one--loop 
scalar diagrams with cubic interactions emerge from \eq{onemast}. Following
\cite{2loop}, we begin by considering the general configuration of a
one--loop one--particle--irreducible diagram with $n$ external legs, 
depicted in Fig.~\ref{Sun}.
Using the matching condition \eq{match3}, we find
\beqa
A^{(1)}_{n,1PI} (p_1, \ldots , p_n) & = & N \, {\rm Tr}(\lambda^{a_1} \cdots 
\lambda^{a_M}) ~\frac{g_3^n}{(4\pi)^{d/2}} (2 \a')^{- d/2} (\a')^n \nl
& \times & \int_{0}^{1}  \frac{dk}{k^2} \,
\left( - \frac{\log k}{2} \right)^{-d/2} \int_{k}^{1} \frac{d z_2}{z_2} 
\int_{k}^{z_2} \frac{d z_3}{z_3} \ldots \int_{k}^{z_{n - 1}} 
\frac{d z_n}{z_n} \nl
& \times & \exp \left[ \sum_{i < j} \left( 2 \a' p_i \cdot p_j 
G^{(1)}(z_i, z_j) \right) \right]~~.
\label{multi3}
\eeqa
As usual, the field theory limit is governed by the multiplier $k$, which 
must be taken to be exponentially suppressed as $\a' \to 0$, so that the 
length of the string loop may become infinite in units of $\a'$. 
The tachyon double pole is regulated, as described in Ref.~(\cite{2loop}), 
by setting $d k/k^2 = \exp (m^2 \a' \log k) ~d k/k$. The appropriate change of 
variables is then
\beqa
k & = & {\rm e}^{- T/\a'} \nl
z_i & = & {\rm e}^{- t_i/\a'}~~.
\label{chvar3}
\eeqa
All puncture coordinates must become exponentially small as $\a' \to 0$, so 
that the correct power of $\a'$ may be generated. 
Defining the Feynman parameters as $x_i = t_{n + 1 - i}/T$, for $i = 1, 
\ldots , n - 1$, we find an expression which may easily be compared with the 
field--theoretic result,
\beqa
A^{(1)}_{n,1PI} (p_1, \ldots , p_n) & = & N \, {\rm Tr}(\lambda^{a_1} \cdots 
\lambda^{a_M}) ~\frac{g_3^n}{(4\pi)^{d/2}} \nl
& \times & \int_0^\infty d T ~T^{n - 1 - d/2} {\rm e}^{- m^2 T} 
\int_{0}^{1} d x_1 \int_{0}^{x_1} d x_2 \ldots \int_{0}^{x_{n - 2}} 
d x_{n - 1} \nl
& \times & \exp \left[ T \sum_{i < j} p_i \cdot p_j 
\left(x_{i j} ( 1 - x_{i j}) \right) \right]~~, 
\label{mul3f}
\eeqa
where $x_{i j} = x_i - x_j$. We note in passing 
that deriving the coefficient of the leading color trace from the Feynman 
rules is not trivial for the general diagram with $n$ legs. In fact, the 
above could be considered as a simple proof that the coefficient is $N$ for 
any $n$. The expression for the integrand as a function of the parameters 
$x_i$ also appears automatically in the most symmetric form; note that this 
is the correct form for arbitrary values of the external momenta $p_i$, 
on-- or off--shell.

Before going on to perform a similar calculation for quartic interactions,
it is worthwhile to pause to consider two instructive special cases of
\eq{mul3f}, and a slight generalization of it in the case of the four--point 
function. First of all we would like to point out that \eq{mul3f} holds in 
field theory for all $n$, including $n = 2$, because of what appears at 
first sight as a fortunate coincidence: in fact in field theory the color 
factor for the two point function is {\em twice} the one appearing in 
\eq{mul3f}, essentially because ${\rm Tr} (\l_a \l_b ) = {\rm Tr} (\l_b 
\l_a )$. However this factor of $2$ is compensated by a symmetry factor of
$1/2$, which is only present for the two--point amplitude. String theory
takes into account these two facts simultaneously and automatically.

Another unexpected feature of the two--point function in field theory is 
the fact that, if one allows the external scalars to take values in the 
$U(1)$ factor of $U(N)$, that is if one allows the indices $a,b$ to take 
the value $0$, one finds that the color factor of the corresponding diagram 
doubles. This is easily seen from Eqs.~(\ref{dun}) and (\ref{dun0}), in the 
Appendix. This fact is unexplained from the point of view of Feynman 
diagrams, but it has a natural explanation in string theory. In fact, if in 
string theory we allow the external legs to be color singlets, the 
amplitude receives a contribution from a {\em new diagram}, the one with 
the two legs inserted on different world sheet boundaries. 
This diagram, which is non--planar from the point of 
view of string theory, contributes at the same order in $N$ because with a 
correctly normalized $U(1)$ generator (see the Appendix) on finds that
${\rm Tr}( \l_0^2 ) {\rm Tr} ( {\bf 1}) = ({\rm Tr}( \l_0 ))^2$. Furthermore,
it is easy to check that the functional form of this new diagram is 
precisely the same as that of the original diagram, with the same overall 
factor and integration region. This is the first simple example of the 
correct handling of non--planar contribution in string theory.
 
We conclude this section by giving a further non--trivial example of a
non--planar contribution to a one--loop amplitude. We consider
the four--point function, which in field theory yields contributions 
proportional to double color traces, such as ${\rm Tr} (\l^{a_1} \l^{a_4})
{\rm Tr} (\l^{a_2} \l^{a_3})$. These double trace contributions naturally 
arise in string theory from the simultaneous insertion of punctures on the 
two different string boundaries. In field theory, on the other hand, these
terms receive contributions from different Feynman diagrams. Choosing for 
example the cyclic order $(1,2,3,4)$ for the external legs, the complete 
color factor in field theory is given by
\beqa
{\cal C}_{1234} & = & N \left[{\rm Tr} \left( \lambda^{a_1} \lambda^{a_2} 
	\lambda^{a_3} \lambda^{a_4} \right)
                          + {\rm Tr} \left( \lambda^{a_4} \lambda^{a_3}
	\lambda^{a_2} \lambda^{a_1} \right) \right] \nonumber \\
                & + & 2 {\rm Tr} \left( \lambda^{a_1} \lambda^{a_2} \right)
                        {\rm Tr} \left( \lambda^{a_3} \lambda^{a_4} \right)
                    + 2 {\rm Tr} \left( \lambda^{a_1} \lambda^{a_3} \right)
                        {\rm Tr} \left( \lambda^{a_2} \lambda^{a_4} \right) 
	\nonumber \\
                & + & 2 {\rm Tr} \left( \lambda^{a_1} \lambda^{a_4} \right)
                        {\rm Tr} \left( \lambda^{a_2} \lambda^{a_3} \right)~~,
\label{colco}
\eeqa
so that the coefficient of the chosen term is $2$. One should keep in  mind, 
however, that the particular color structure ${\rm Tr} (\lambda^{a_1} 
\lambda^{a_4}) {\rm Tr} (\lambda^{a_2} \lambda^{a_3})$ receives contributions 
also from two other distinct Feynman diagrams, corresponding to the 
non--cyclic permutations of the external particles in the original one (in 
the present case, the orderings $(1,3,2,4)$ and $(1,4,2,3)$). 
String theory must, and does, assemble the contributions 
of the different diagrams to the chosen color 
structure into a single string configuration. This can be verified by using 
the non-planar Green function \eq{gr1np}, and considering all possible ways 
of inserting the punctures on the boundaries. Since we wish to place the 
puncture $z_4$ on the same boundary as $z_1 = 1$, it must lie on the positive 
real axis in Fig.~\ref{sch1}. As explained above, this leads to the 
integration region $k \leq z_4 \leq 1$. For the other two punctures, on the 
negative axis, the integration region is $ - 1 \leq \{z_2, z_3\} \leq - k $, 
with no restriction on the relative ordering of $z_2$ and $z_3$. 
Changing variables to $x_2 = - z_2$ and $x_3 = - z_3$, we have once again
four punctures on the positive real axis, which can be placed in the interval
$[k,z_1 = 1]$ in six different orderings. Note that the six orderings are 
distinguishable in the field theory limit, because the logarithmic term in the
generic Green function has a different field theory limit depending on the 
ordering, $\log(z_i \pm z_j) \to \log(z_i)$, for $z_i > z_j$. An explicit 
calculation shows that the orderings $x_2 \leq x_3 \leq z_4$ and
$z_4 \leq x_3 \leq x_2$ conspire to reconstruct the contribution to the 
chosen trace of the field theory diagram with cyclic ordering $(1,2,3,4)$, 
with the correct overall factor of $2$. Similarly, the orderings $x_2 \leq z_4 
\leq x_3$ and $x_3 \leq z_4 \leq x_2$ reconstruct the contribution of the 
diagram with cyclic ordering $(1,2,4,3)$, while the remaining two orderings
give the last diagram. Once again, the building blocks of the final result 
are assembled in novel and non--trivial way.

\subsection{One--loop quartic interactions}
\label{olq}

\begin{figure}[ht] 
\begin{center} 
{\scalebox{0.5}{\includegraphics{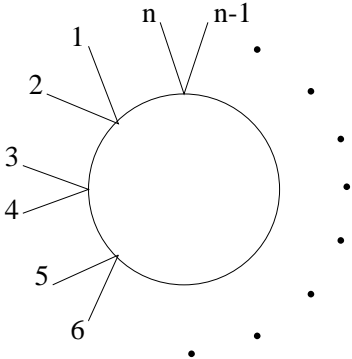}}}
\end{center} \caption{Multi--leg one--loop one--particle--irreducible diagram
in a $\Phi^4$ theory.}
\label{Doublesun}
\end{figure}

As in the previous section, it is possible to give a general expression for
the one--loop, color--ordered, one--particle--irreducible diagram with $n$
external legs, but only quartic vertices, shown in Fig.~\ref{Doublesun}.
Using the matching condition \eq{match4} in \eq{onemast}, we get an 
expression very similar to \eq{multi3},
\beqa
A^{(1)}_{n,1PI} (p_1, \ldots , p_n) & = & N \, {\rm Tr}(\lambda^{a_1} \cdots 
\lambda^{a_M}) ~\frac{g_4^{n/2}}{(4\pi)^{d/2}} (2 \a')^{n/2 - d/2} \nl
& \times & \int_{0}^{1}  \frac{dk}{k^2} \,
\left( - \frac{\log k}{2} \right)^{-d/2} \int_{k}^{1} \frac{d z_2}{z_2} 
\int_{k}^{z_2} \frac{d z_3}{z_3} \ldots \int_{k}^{z_{n - 1}} 
\frac{d z_n}{z_n} \nl
& \times & \exp \left[ \sum_{i < j} \left( 2 \a' p_i \cdot p_j 
G^{(1)}(z_i, z_j) \right) \right]~~.
\label{multi4}
\eeqa
It is clear, however, that in this case we must introduce proper times for 
only one half of the integration variables, in order to compensate for the 
overall factor of $(\a')^{n/2}$. To obtain the configuration shown in 
Fig.~\ref{Doublesun}, the appropriate change of variables is
\beqa
z_i & = & {\rm e}^{- t_i/\a'}~~, \qquad \qquad \, ({\rm i ~~odd}) \nl
z_i & = & y_{i} ~z_{i - 1}~~, \qquad \qquad   ({\rm i ~~even})
\label{var4} \\
k   & = & {\rm e}^{- T/\a'}~~, \nonumber
\eeqa
with no proper time associated to the variables $y_i$ and, as usual,
$z_1 = 1$. Using \eq{var4} and neglecting terms suppressed as 
$\a' \to 0$, one gets
\beqa
A^{(1)}_{n,1PI} (p_1, \ldots , p_n) & = & N ~{\rm Tr}(\lambda^{a_1} \cdots 
\lambda^{a_M}) ~\frac{(2 ~g_4)^{n/2}}{(4\pi)^{d/2}} \label{mul4} \\
& \times & \int_0^\infty d T ~T^{- d/2} {\rm e}^{- m^2 T} 
\int_0^1 \frac{d y_2}{y_2} \int_0^T dt_3 \ldots \int_0^{t_{n - 3}} 
d t_{n - 1} \int_0^1 \frac{d y_n}{y_n} \nl
& \times & \exp \left[ \sum_{i < j ~{\rm odd}} (p_i + p_{i + 1})\cdot 
(p_j + p_{j + 1}) \left(t_j - t_i - \frac{(t_j - t_i)^2}{T} \right) \right]
\, . \nonumber
\eeqa
The integrals in $d y_i$, ranging over finite regions of moduli space, are 
logarithmically divergent and need to be regularized. This divergence is 
actually a further manifestation of the tachyonic nature of the bosonic 
string ground state, and must be dealt with by the same method used to handle
the double pole $d k/k^2$: one must substitute $y_i^{-1} \to \exp(m^2 \a' 
\log y_i)$. In this case, however, there is no proper time associated with 
$y_i$ so the exponential factors are suppressed as $\a' \to 0$ and must be 
neglected. The product of all $d y_i$ integrals then yields simply a factor 
of unity. This procedure can be further justified by noting that the factors 
of $y_i^{-1}$ arise from the factors $(V_i'(0))^{-1}$ in \eq{onemast}, which 
are characteristic of tachyon propagation and are absent, for example, in
gluon amplitudes. In fact \eq{onemast} can be consistently continued to an 
arbitrary value of the string intercept $a = - m^2 \a' \neq 1$, as was done 
in~\cite{aalo}. In that case one finds that the measure of integration changes 
as $d k/(k^2 \prod_i V_i'(0)) \to (d k/k) (k \prod_i V_i'(0))^{-a}$, 
indicating that the singularities generated by the projective transformations 
$V_i'(0)$ are of the same type as the double pole in the multiplier $k$.

With this regularization and changing variables to $x_i = 1 - t_i/T$, we find
the simple formula
\beqa
A^{(1)}_{n,1PI} (p_1, \ldots , p_n) & = & 2^{n/2} ~N ~{\rm Tr}(\lambda^{a_1} 
\cdots \lambda^{a_M}) ~\frac{g_4^{n/2}}{(4\pi)^{d/2}} \label{mul4f} \\
& \times & \int_0^\infty d T ~T^{(n - 2 - d)/2} {\rm e}^{- m^2 T} 
\int_0^1 dx_3 \int_0^{x_3} d x_5 \ldots \int_0^{x_{n - 3}} 
d x_{n - 1} \nl
& \times & \exp \left[ T \sum_{i < j ~{\rm odd}} (p_i + p_{i + 1})\cdot 
(p_j + p_{j + 1}) \left(x_{i j} (1 - x_{i j}) \right) \right]
\, . \nonumber
\eeqa
Once again, this matches the result arising from the Feynman diagram in 
Fig.~\ref{Doublesun}, with the restriction that
paired external legs must not be of the same color. This provides a simple
proof of the fact that for such diagrams the coefficient of the
leading color trace is $2^{n/2} N$. 

We conclude this section by noting a special case that arises 
in the computation of these loop amplitudes, when two consecutive external
particles ($(2n-1,2n)$) annihilate in two colorless states running in
the loop.
Let us, for instance, focus on the special case of \eq{mul4f} where $n
= 2$; this of course represents a tadpole diagram. For this diagram
\eq{mul4f} yields 
\beq
A^{(1)}_{2,1PI} (p_1, p_2) = 2 ~N ~{\rm Tr}(\lambda^a 
\lambda^b) ~\frac{g_4}{(4\pi)^{d/2}} \int_0^\infty d T ~T^{- d/2}
{\rm e}^{- m^2 T}~~,
\label{tad4}
\eeq
which is one half of the result obtained in field theory. This discrepancy
can be understood by observing that the color factor in field theory 
is of the form
\beq
{\cal C}_{a b} = 8 ~N {\rm Tr} (\l_a \l_b) = d_{a b \mu} 
d^{\mu \c}_{\; \; \c} + 2 d_{a \mu \nu} d_b^{\; \mu \nu}~~.
\label{tadcol}
\eeq
The first contraction of $d$--tensors, which contributes one half of the 
total result, represents an `anomalous' color flow corresponding to a
$\Phi^3$ tadpole diagram, in which the colored scalars $a$ and $b = a$ 
annihilate into a color singlet state which then self--interacts. Clearly in 
this channel the full $U(N)$ color flow is prohibited, and only the $U(1)$ 
factor contributes. This term is missed by \eq{mul4f}, but can be
reproduced by first generating a one--particle--reducible $\Phi^3$
diagram and then deleting the zero--momentum propagator, thus attaching
the loop to the external legs. A similar peculiarity will arise when
we will consider the two--loop $\Phi^4$ vacuum bubbles. Also in that case a 
tadpole--like configuration forcing the color flow to 
be restricted to $U(1)$ is present. This kind of term is always
related to corners of moduli space corresponding
to one--particle--reducible diagrams, and this is signaled by a color
factor which displays a combinations of $d_{\a \b \c}$ symbols typical of
these diagrams.
 

\sect{Two--loop diagrams}
\label{twoloop}
\vskip 0.5cm

\begin{figure}[ht] 
\begin{center} 
{\scalebox{0.5}[0.7]{\includegraphics{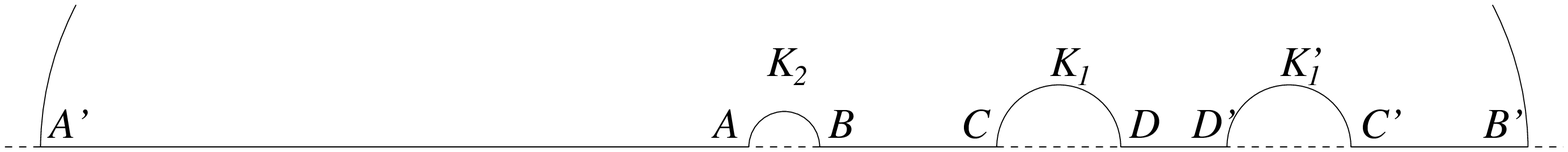}}}
\end{center} \caption{In the Schottky parametrization, the two--annulus
corresponds to the part of the upper--half plane which is inside the big
circle passing through $A'$ and $B'$, and which is outside the
circles $K_1$, $K_1'$ and $K_2$.}
\label{2lf}
\end{figure}

Generalizing the approach described in the previous
section to the two--loop case is not a straightforward task. 
First, the explicit expressions of  the measure and of 
the other geometrical objects present in \eq{hmastac} are 
more complex, so that the computation of physically 
interesting quantities (such as Yang--Mills amplitudes)
has to be performed by means of a computer program. 
Second, there are also conceptual novelties, and the procedure 
described in the previous section cannot be applied directly to a 
two--loop calculation. Many of the new features are related to 
the fact that now the string world--sheet is a two--annulus and 
some of the simplifying choices that are usually made at one--loop
are not possible any more. For example, in Eq.~(\ref{onemast})
the fixed points played no role, because they could be gauge--fixed 
to zero and infinity. In two--loop calculations, on the other hand, 
the shape of the string world--sheet can vary in a non--trivial manner. 
In fact, the measure \eq{hmeasure} crucially depends on at least one 
of the fixed points, which means that the relative position of the 
two holes cannot be fixed. Thus, for a better understanding of the 
new geometrical features, it is worthwhile to start the study of 
two--loop string amplitudes by considering in detail the Schottky
parametrization, as it  arises from the sewing procedure leading 
to \eq{hmastac}. Basically, the two--loop surface can be constructed 
starting at one loop and identifying two external legs. This is done by
cutting away from the one--loop string world--sheet two circles, and
identifying their boundaries. If one chooses to sew together the puncture 
fixed at $z_1 = 1$ with one of the other legs on the same boundary 
({\em i.e.} with $z_i >0$ in order to construct a planar diagram), 
one obtains the two--loop surface depicted in \fig{2lf}.

We fix projective invariance by choosing $\eta_2 = 0$, $\xi_2 \to \infty$
and $\xi_1 = 1$. Then the positions and radii of the circles in \fig{2lf}
are completely determined as functions of the multipliers $k_1$, $k_2$ and
of the fixed point $\eta_1$; in fact, following~\cite{2loop}, one can 
verify that
\beq \label{abcd}
B = - A = \sqrt{k_2}~~,~~~ 
C = \frac{\eta_1 - \sqrt{k_1}}{1 - \sqrt{k_1}}~~,~~~
D = \frac{\eta_1 + \sqrt{k_1}}{1 + \sqrt{k_1}}~,
\eeq
\beq
D' = \frac{1 + \eta_1 \sqrt{k_1}}{1 + \sqrt{k_1}}~~,~~~
C' = \frac{1 - \eta_1 \sqrt{k_1}}{1 - \sqrt{k_1}}~~,~~~
B' = - A' = \frac{1}{\sqrt{k_2}}~~. \nonumber
\eeq
One can check that the points $A$, $B$, $C$ and $D$ are 
identified with $A'$, $B'$, $C'$ and $D'$, respectively, under 
the action of the two generators of the two--loop Schottky group, 
{\it i.e.} the projective transformations mapping the circles $K_{1,2}$
into their images $K_{1,2}'$.
\vspace{.2cm}
By cutting open a two--annulus, it is possible to map the two segments 
$(AA')$ and $(DD')$ onto the inner boundaries of the
world--sheet, which is natural since their length depends only on $k_1$
and $k_2$. Then the union of $(BC)$ and $(C'B')$ represents the external 
boundary. Note that, in order to avoid a degenerate surface, the various
identified circles should not overlap. This simple consideration
gives a first constraint on the region of integration of the string
moduli. In fact, by requiring that $B$ does not touch $C$ one obtains
\beq
\eta ~> \sqrt{k_1} + \sqrt{k_2} - \sqrt{k_1 k_2}~~,
\label{eta}
\eeq
while the same requirement on the segment $(DD')$ leads to
\beq
D' - D = \frac{1 - \sqrt{k_1}}{1 + \sqrt{k_1}} ~(1 - \eta_1)~~,
\label{ometa}
\eeq
so that $\eta < 1$. Now the interpretation of the three moduli,
$k_1$, $k_2$ and $\eta_1$, is particularly simple. In fact,
$\sqrt{k_2}$ is the radius of the circle $K_2$, while the radii of
$K_1$ and $K_1'$ are equal and depend on $\sqrt{k_1}$
\beq
R_{K_1} = \sqrt{k_1} \left({1- \eta_1\over 1-k_1}\right)~~.
\label{rk1}
\eeq
Furthermore, $\eta_1$ turns out to be inside $K_1$, while the point 
$\xi_1=1$ is inside $K_1'$. Therefore, in this configuration, the  
circle $K_1'$ is fixed while $K_1$ can move, depending on the value 
of $\eta_1$. In particular, if the point $D'$ is very close to $D$, 
$\eta_1$ is almost equal to $1$, while if $C$ is near to $B$, then 
$\eta_1$ is of order of $\sqrt{k_1} + \sqrt{k_2}$. In other words, it 
is possible to interpret $\eta_1$ as the ``distance'' between the two 
loops. When $\eta_1 \to 1$, one may expect the field theory limit (at 
least for cubic interactions) to yield reducible diagrams with the two 
loops widely separated, while when $\eta_1 \to \sqrt{k_1} + \sqrt{k_2} 
\to 0$ one should obtain the irreducible diagrams with the two loops 
attached to each other.  

Another observation relevant for deriving the region of integration of 
world--sheet moduli is that in the explicit form of all geometrical objects
(for instance, the Green function or the measure) the multipliers $k_1$
and $k_2$ appear symmetrically, reflecting the equivalence of the two loops. 
Therefore, in order to avoid double counting of equivalent configurations, one
can order the multipliers, by choosing for example $k_2 < k_1$. 
Note that the multipliers will always be associated with field theory 
proper times, so in the zero--slope limit their ordering should always be 
interpreted as strong ordering. 

A final remark has to be made about the possibility of taking the
attractive fixed point $\eta$ bigger in modulus than the repulsive one
$\xi$. As was shown in~\cite{kajun}, in the closed string case these
configurations are related to the one with $|\eta|<|\xi|$ by the
residual part of the modular invariance group which survives in the
field theory limit. In the open string case there is no modular
invariance, but these surfaces should describe the propagation of
closed string states, and they should not be included in our analysis.
Thus, for our purposes, we will always restrict $|\eta|$ to be less than 
$|\xi|$. Summarizing, we have found for 
$k_1$, $k_2$ and $\eta_1$, in the field theory limit, the same region of 
integration derived by Roland~\cite{kajun} for the closed string, {\it i.e.} 
$0 \leq \sqrt{k_2} < \sqrt{k_1} < |\eta_1| \leq 1$. We are now ready to move 
on to the evaluation of two--loop diagrams, starting with the simplest 
ones to verify our assumptions.

\subsection{Vacuum bubbles}

Let us start by briefly describing the simplest amplitude,
the two--loop vacuum bubble with cubic interaction.
In this case ($M = 0$ and $h = 2$), with the projective gauge choice
and integration region described above, and using the matching condition
for cubic interaction , \eq{match3}, \eq{hmastac} simply becomes
\beqa
A^{(2)}_0 & = & \frac{N^3}{(4\pi)^d} \, \frac{g^{2}_3}{16} \,
(2 \a' )^{3-d} \int_0^1 \frac{d \eta_1 }{(1 -  \eta_1)^2} \int_0^{\eta_1} 
\frac{d k_1}{k_{1}^{2}} \int_0^{k_1} \frac{d k_2}{k_{2}^{2}} 
\nonumber \\
& \times & \left[\frac{1}{4} \left(\log k_1 \log k_2 -
\log^2 \eta_1 \right) \right]^{-d/2}~~, 
\label{vacuums}
\eeqa
In \eq{vacuums} only the determinant of the period matrix is needed; at 
leading order in the two multipliers, it is
\beq
\det ( - i \tau_{\mu \nu} ) = \frac{1}{4 \pi^2} \left[ \log k_1
\log k_2 - \log^2 \eta  \right]~.
\label{permat}
\eeq
Note that, as expected, the two multipliers $k_1$ and $k_2$
play the same role and all the expressions are
symmetrical in the exchange of $k_1$ and $k_2$.

As discussed in the previous section, we expect a contribution
to the field theory result from the limit $\eta_1\rightarrow 1$
(together with ${k_1,k_2} \to 0$); in this case, the appropriate change 
of variables is
\beq
t_1 = - \a' \log k_1 \,, ~~~ 
t_2 = - \a' \log k_2 \,, ~~~ 
t_3 = - \a' \log (1 - \eta_1)~~.
\label{ti}
\eeq
Introducing the mass $m^2$ to regulate quadratic poles in the usual way,
and neglecting terms suppressed as $\a' \to 0$,
\eq{vacuums} in this region yields
\beq
A^{(2)}_{0,{\rm red}} = \frac{N^3}{(4\pi)^d}\,\frac{g^{2}_3}{2} \,
\int_0^{\infty} d t_3 \int_0^{\infty} d t_2 \int_0^{t_2} d t_1~
{\rm e}^{- m^2 (t_1 + t_2 + t_3)} ~(t_1 \, t_2)^{-d/2}~~.
\label{vacuumr}
\eeq
Since \eq{vacuumr} is symmetrical in $t_1$ and $t_2$, it is possible to 
perform the integration over $t_1$ and $t_2$ independently 
from $0$ to $\infty$ by introducing a factor of $1/2$. In this way one 
obtains the same result of the reducible vacuum bubble of the $\Phi^3$ 
field theory defined by \eq{lag3}, including the correct normalization.
Again, the factor of $1/4$ in the normalization of this diagram in field 
theory is a combination of color and symmetry factors, which are unified 
in the present approach.

The second expected contribution comes from the limit $\eta_1\to 0$ and 
should give the irreducible vacuum diagram. In this case each loop is made of
two different propagators, and this suggests a different identification between
field theory proper times and string variables. In fact, the experience 
acquired at one loop leads us to expect that the multipliers $k_i$ should be 
associated with the length of an entire loop. Thus we set
\beq
q_1 = {k_2\over\eta_1} = {\rm e}^{- t_1/\a'}~,~~
q_2 = {k_1\over\eta_1} = {\rm e}^{- t_2/\a'}~,~~
q_3 = \eta_1 = {\rm e}^{- t_3/\a'}~.
\label{qvariable}
\eeq
With this choice, \eq{vacuums} becomes
\beqa
A^{(2)}_{0,{\rm irr}} & = & \frac{N^3}{(4\pi)^d} \, \frac{g^{2}_3}{2}
\, \int_0^{\infty} d t_3 \int_0^{t_3} d t_2 \int_0^{t_2} d t_1~
{\rm e}^{- m^2 (t_1 + t_2 + t_3)} \nl
& \times & (t_1 t_2 + t_1 t_3 + t_2 t_3)^{-d/2}~~.
\label{vacuumir}
\eeqa
Since \eq{vacuumir} is completely symmetrical, we can introduce a factor of
$1/3!$ and perform the three integrals independently from 0 to $\infty$.
In this way one correctly reproduces the irreducible vacuum bubble of
our $\Phi^3$ theory. Note that by using a single starting formula, 
\eq{vacuums}, we have been able to obtain two diagrams which have a 
different weight. While in field theory this relative factor of $3$ 
between the two vacuum bubble amplitudes is due to the 
combination of different combinatorial and color factors, 
in the string approach this relative 
normalization appears because \eq{vacuums} has different symmetry
properties in the two regions of moduli space that yield
the two vacuum bubbles.

It is interesting to see what happens if one considers also 
non--planar string world--sheets. From \fig{2lf}, one can see that a
non--planar surface may arise if the circle $K_1$ is centered on the 
negative axis. The identifications established in the planar case still 
hold and by following them it is easy to realize that the surface has 
only one boundary. When $\eta$ approaches to zero, the calculations
follows exactly the same pattern we have just seen and the result is
again \eq{vacuumir}, but with a factor of $N$ only, instead
of $N^3$, since here we have only one boundary. If however one
tries to mimic the limit which gave the reducible bubble, that is
$|\eta|\to 1$, one sees that there is no singularity in \eq{vacuums}.
In fact, now $\eta$ approaches $-1$ and it is not possible to
associate a Schwinger proper time to the combination $(1-\eta)$:
this corner of moduli space gives a vanishing contribution showing that 
only the irreducible bubble receives subleading corrections in $N$. 
By using the equations given in Appendix, it is easy to recover the same
result from a field theory analysis.

We conclude our discussion of two--loop vacuum bubbles by briefly considering
the single diagram arising in the case of quartic interaction. Here we
see at work the mechanism that was suggested for the one--loop $\Phi^4$
tadpole, in \secn{oneloop}. In fact, the field theory color factor of the 
vacuum bubble is of the form
\beqa
{\cal C} & = & d^\a_{\; \, \a \mu} \,d^{\mu \c}_{\; \; \c} + 
2 \, d_{\a \c \mu} d^{\a \c \mu} \nl
& = & 2 \, N^3 + 2 \, N \, (N^2 + 1)~~,
\label{colf}
\eeqa
where we emphasized the different origin of the various factors of $N$.
The symmetry factor, on the other hand, is $1/8$. Using string theory, and
the matching condition \eq{match4}, one sees that the overall 
normalization of this diagram in the planar case is $N^3 g_4$.
Now, however, one must introduce proper times only for the two multipliers,
since there are only two propagators in the diagram. Thus the overall 
symmetry factor for the completion of the integration region is $1/2$ 
instead of $1/6$, for both color flow patterns. If we proposed to add them, it
would appear that the string gives a result too big by  factor of two.
The solution to this puzzle lies in the observation that in both 
configurations one needs to integrate over a finite range in $\eta$,
extending to $\eta \to 1$, as in \eq{vacuums}. One must regulate the 
singularity as $\eta \to 1$, without having a proper time associated with 
$1 - \eta$. Singularities of this kind were studied in Refs.~(\cite{big,maru}),
and the correct prescription (a $\zeta$ function regularization) turns out 
to be that these integrals yield precisely a factor of $1/2$.
Notice that the non--planar contribution can arise only from the 
irreducible diagram, which justifies the fact that the coefficient 
of $N$ in \eq{colf} is $1/2$ of the coefficient of $N^3$.

\subsection{Two--point amplitudes}

\begin{figure}[ht] 
\begin{center} 
{\scalebox{0.55}{\includegraphics{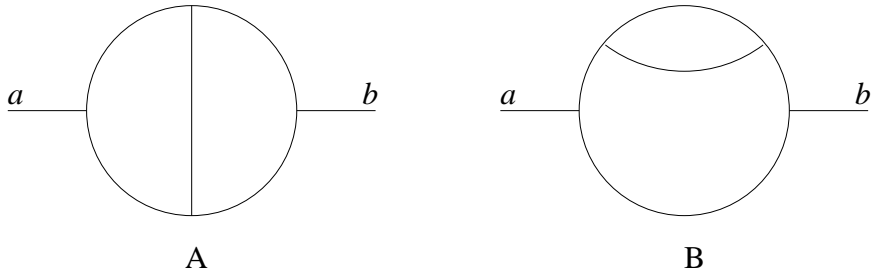}}}
\end{center}
 \caption{Irreducible two--loop diagrams contributing to the
two--point function of the $\Phi^3$ theory.}
\label{2l2pf}
\end{figure}

In this section, scalar amplitudes with two external states are considered;
in particular, the irreducible diagrams of Fig.~\ref{2l2pf} are derived
directly from \eq{hmastac}, without using the simplified procedure described 
in \cite{2loop}. Instead, we follow the procedure just employed for the
vacuum bubbles and, for each diagram, we look at the appropriate corners 
of the integration region of the various string variables. 
The new ingredients needed in \eq{hmastac} in presence of external 
particles are the two--loop bosonic Green function and, if one wants
to extrapolate the result off--shell, the new expressions of the local 
coordinates $V_i'(0)$. It was shown in \cite{2loop,rolsat1} that from 
the Green function in \eq{Green} the general structure of two--loop 
$\Phi^3$ diagrams can be recovered. Moreover Roland and Sato showed 
in \cite{rolsat1} that the string master formula (\ref{hmastac}) reduces 
to the particle theory amplitudes in the world--line
approach, even in the multi--loop case. However, in those papers there 
is no derivation for the region of integration over the punctures $z_i$;
this should be given in terms of the parameters determining the shape
of two--loop world--sheet. Lacking this information, it
is difficult to fix the correct normalizations of the various
diagrams, since different corners of the region of integration over the
punctures can contribute to the same field theory diagram.
We will now show that, in order to determine the integration region
for the punctures $z_i$, it is sufficient to use the simple geometrical 
description of \fig{2lf}, and the analysis of vacuum amplitudes 
outlined in the previous sections. In fact, as discussed in the Introduction,
a general feature of the string approach is that the calculations of 
diagrams with different number of external legs are closely related 
to each other. For instance, it is 
clear from the analysis of vacuum bubbles that, in order to construct 
irreducible diagrams, the limit $\eta_1\rightarrow 0$ must be considered. 
At this point one can freeze the world--sheet shape and put the punctures 
in all possible configurations on the three boundaries, remembering that, 
since we are interested in planar amplitudes, both states must 
lie on the same boundary of the two--annulus. 

Let us start by considering the case in which the punctures are inserted 
on the internal boundary represented by the segment $(AA')$. They should then
be integrated in the interval $[-1/\sqrt{k_2},\sqrt{k_2}]$. Analogously,
if they are on the other internal boundary, the region of integration is
the interval $[D,D']$, while for the contributions of the external 
boundary each $z_i$ must be allowed to range between $B$ and $C$, and 
between $C'$ and $B'$. In the configuration with small multipliers 
($q_i \rightarrow 0$) the world--sheet becomes a graph, and each boundary 
degenerates into the union of two distinct field theory propagators. 
It is interesting to note that it is possible to identify the specific 
corners of moduli space associated with each first quantized field 
theory diagram, already at the string level. In fact, if $z \in [A',A]$, 
the contributions obtained correspond to a diagram with the external 
particles always attached to the first loop; but it turns out that they are 
emitted from the propagator shared by the two loops if $z\in[-1,-\eta_1]$, 
while the intervals $z\in[-\eta_1,A]$ and $z\in[A',-1]$ correspond to 
emission from the propagator not shared with the second loop.

In order to show that this identification is correct, let us calculate 
the two--loop diagram with one external state in the region
$[-1,-\eta_1]$ and the other in the region $[A',-1]$ that should 
contribute to the first diagram of \fig{2l2pf}. As for the local coordinates,
here we impose again that $V_i'(0) = |z_i|$, since the punctures
are on a boundary that, from the point of view of the Schottky
parametrization, is identical to the one encountered in the one--loop
case. Of course, other choices lead to the correct result
\cite{2loop,rolsat2} and only a Yang-Mills calculation, where higher--order 
terms in $q_i$ are relevant, can discriminate among the various options. 
In this region, the variable $U$ of \eq{STU} can be approximated as 
$U\sim |z_1|^{-1}$. Thus from \eq{hmastac} one can derive a first 
contribution to the diagram in \fig{2l2pf}, by treating the quadratic 
poles of the $q_i$ variables in the usual way, and by introducing proper times
$t_i$ as in \eq{qvariable}. One finds
\beqa
A^{(2)}_a (p_1, p_2) & = &
\frac{N^2}{(4\pi)^{d}}\frac{g^4_3}{4} \d^{ab} \int_0^\infty d t_2 
\int_0^{t_2} d t_1 \int_0^{t_1} d t_3
\int_0^{{1\over 2}(t_2 + t_3)} d t_{z_1}
\int_0^{t_3} d t_{z_2} \nl & \times & {\rm e}^{-m^2 (t_1 + t_2 + t_3)}~
\Delta^{-d/2} \exp{\left[p_1 \cdot p_2 \, 
G_a (t_1, t_2, t_3, t_{z_1}, t_{z_2})
\right] }~~,
\label{2p1}
\eeqa
where the Green function for this diagram is given by
\beq
G_a (t_1, t_2, t_3, t_{z_1}, t_{z_2}) = t_{z_1} + t_{z_2} - \Delta^{-1} \left(
(t_1 + t_2) t_{z_2}^2 + (t_1 + t_3) t_{z_1}^2 + 
2 t_{z_1} t_{z_2} t_1 \right) \, ,
\label{gapple}
\eeq
and where $\Delta = (t_1 t_2 + t_1  t_3 + t_2 t_3)$. The Schwinger proper
times $t_{z_i}$ are related, as expected, to the positions of the 
punctures by $t_{z_i} = - \a' \log|z_i|$.
It can be checked that the integrand of \eq{2p1} is exactly what one 
expects from a field theory calculation, but the region of integration 
over proper times appears strange. In fact, the whole expression is not 
symmetric in the exchange of any two proper times, unlike the case of vacuum 
diagrams, \eq{vacuumir}. Thus, it is not possible 
to immediately complete the integration region of $t_1$ and $t_3$ extending
it to infinity; moreover, from the field theory analysis one would expect 
the proper time $t_{z_1}$ to vary between $0$ and $t_2$, while the string 
result covers only part of this region.

\begin{figure}[ht] 
\begin{center} 

\rotatebox{270}{\scalebox{0.7}{\includegraphics{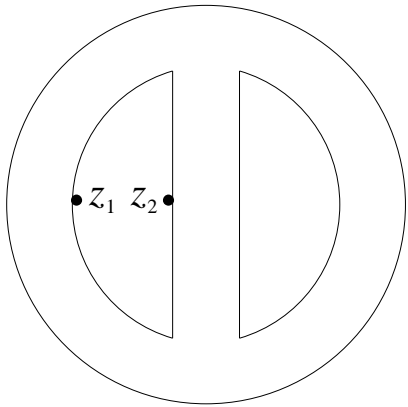}}}
\end{center}
\caption{The world--sheet representation of the two--loop 
contribution derived in the text.}
\label{strrap}
\end{figure}

These problems are treated by taking into account also the other regions
of the integration over the punctures that contribute to the field theory 
diagram of \fig{2lf}. For instance, another configuration that should be 
considered is $z_2\in [-\eta,-\sqrt{k_2}]$ with $z_1$ remaining in the same 
interval as before. In this case $U$ can be approximated as $U \sim 
\eta/|z_2|$, and from \eq{hmastac} one gets
\beqa
A^{(2)}_b & = &
\frac{N^2}{(4 \pi)^{d}} \frac{g^4_3}{4} \d^{ab} \int_0^\infty d t_2 
\int_0^{t_2} d t_1 \int_0^{t_1} d t_3
\int_{t_3}^{{1\over 2}(t_2 + t_3)} d t_{z_1} 
\int_0^{t_3} d t_{z_2} \,{\rm e}^{- m^2(t_1 + t_2 + t_3)} \nl
& \times & \Delta^{-d/2} \exp{\left[ p_1\cdot p_2 \, 
G_b (t_1, t_2, t_3, t_{z_1}, t_{z_2})
\right] }~~,
\label{2p1b}
\eeqa
where
\beqa
G_b (t_1, t_2, t_3, t_{z_1}, t_{z_2}) & = & t_{z_1} - t_{z_2} - \Delta^{-1}
\left[ (t_2 + t_3) (t_3 - t_{z_2})^2 \right. \label{gappleb} \\
& + & \left. (t_1 + t_3) (t_{z_1} - t_{z_2})^2 - 
2 (t_{z_2} - t_{z_1}) (t_{z_2} - t_3) t_3 \right]~.
\nonumber
\eeqa
These results seem completely different from \eq{2p1}, and in fact it
is difficult to relate them to the diagram of \fig{2lf}, since the 
Green function is not the one expected. However, this happens just because 
\eq{2p1b} is not expressed in terms of the most convenient variables; 
in fact, by changing $t_{z_1} \to t_2 + t_3 - t_{z_1}$ the integrand in 
\eq{2p1b} becomes exactly that of \eq{2p1} and the region of integration
over $t_{z_1}$ becomes the interval $[{1\over 2}(t_2 + t_3),t_2]$. 
At this point the two contributions can be summed by simply  
extending the integration of $t_{z_1}$ from $0$ to $t_2$ and, as far as
the punctures are concerned, the expected field theory 
result is obtained. Note that the integrand of \eq{2p1} is symmetric
under the simultaneous exchanges $t_{z_1} \leftrightarrow t_{z_2}$ 
$t_2 \leftrightarrow t_3$, as implied by the world--sheet representation of 
these regions of integration (see \fig{strrap}).
This consideration suggests that the symmetry in the exchange among $t_1$, 
$t_2$ and $t_3$ must be obtained only when the external states are put on 
the other two boundaries. In fact, when the two punctures are on the 
boundary $(D,D')$ the final result is exactly that of \eq{2p1}, but with 
the roles of $t_2$ and $t_1$ exchanged. 
Some changes should however be introduced into the 
procedure, since this boundary is not in the usual ``one--loop form'' (that
is with the multipliers at $0$ and $\infty$). First, the local
coordinates should be expressed in terms of the abelian differential of 
this loop; second, the variables $z_i$ are not the ones that are directly
related to the Schwinger proper times. In fact, the minimum value of
$z$ in this configuration is $D$, \eq{abcd}, and not $\sqrt{k_1}$, as 
one should expect from the symmetry with the case just considered. 
In this case the string variable directly related to the Schwinger 
proper time is
\beq
x = \left( \frac{z - \eta}{z - 1} \right)~,
\label{xvar}
\eeq
which ranges in the interval $[- {1\over\sqrt{k_1}}, -\sqrt{k_1}]$, as 
expected by symmetry. With these changes, the analysis of the integration 
on the second boundary is completely similar to the one discussed above. 

The contributions coming from the configurations with the punctures on 
the external boundary can be calculated along the same line. For the local
coordinate one can choose the most symmetric combination of the two 
abelian differentials, since this boundary surrounds both loops
\beq
V_i'(0)^{-1} = \omega_1 +\omega_2~.
\eeq
The complete result will now be the combination of four regions, corresponding
to $z_1 \to B, B'$ and $z_2 \to C, C'$. The four regions combine into a 
single integral, as for the two contributions (\ref{2p1}) and (\ref{2p1b}), 
and give the same result as \eq{2p1} with $t_3$ and $t_1$ exchanged.

Taking into account all boundaries, one gets a completely symmetric
expression, that should be further multiplied by a factor of two, since 
in all the computations presented the role of $z_1$ and $z_2$ can be 
exchanged, as was discussed after \eq{hmeasure}. At this point one
can perform the integration over $t_1$, $t_2$ and $t_3$ independently,
introducing, a factor of $1/3!$ that cancels the factor of two coming 
from the $z_1 \leftrightarrow z_2$ symmetry, and the triple counting
arising from the three different first quantized diagrams. Thus the complete
contribution for the field theory configuration of \fig{2lf} is simply
\beqa
A^{(2)}_a (p_1, p_2) & = &
\frac{N^2}{(4\pi)^{d}}\frac{g^4_3}{4} \d^{ab} \int_0^\infty d t_1
\int_0^{\infty} d t_2 \int_0^{\infty} d t_3
\int_0^{t_2} d t_{z_1}
\int_0^{t_3} d t_{z_2} \, {\rm e}^{-m^2 (t_1 + t_2 + t_3)} \nl
& \times & \Delta^{-d/2} \exp{\left[p_1 \cdot p_2 \, 
G_a (t_1, t_2, t_3, t_{z_1}, t_{z_2})
\right] }~~,
\label{2pfin}
\eeqa
the expected result.

\begin{figure}[ht] 
\begin{center} 
\rotatebox{270}{\scalebox{0.7}{\includegraphics{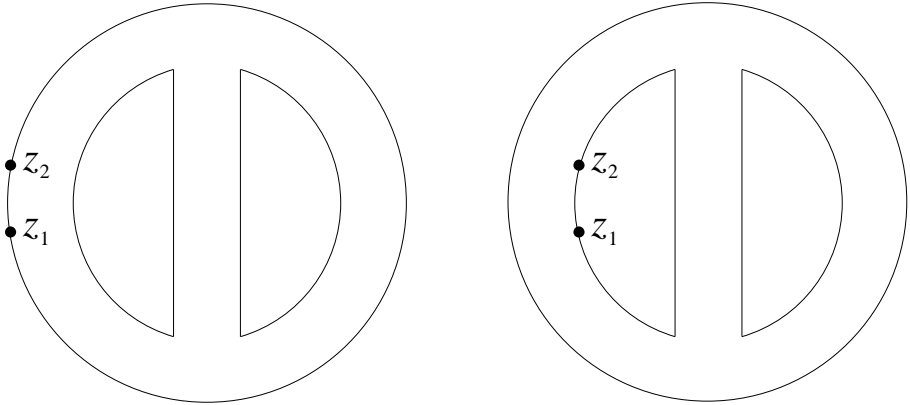}}}
\end{center}
\caption{Two different world--sheet configurations that contribute to the 
same first quantized diagram}
\label{fish}
\end{figure}

By using the identification among field theory propagators and regions of
integration, it is possible to derive also, in a similar way, the other 
irreducible two--point $\Phi^3$ diagram in Fig.~(\ref{2l2pf}). This diagram 
has different symmetries, which implies, in field theory, that its 
normalization differs from the one of \eq{2pfin} by a factor of two. Without 
entering the details of the calculations, it is easy to see the string 
origin of this difference. In fact, as should be clear from \fig{fish}, in 
this case it is possible to consider two world--sheet configurations for
each first quantized diagram, besides the usual exchange between the
external legs $z_1$ and $z_2$; this fact is eventually responsible
for the different normalization. The result is
\beqa
A^{(2)}_c (p_1, p_2) & = &
\frac{N^2}{(4\pi)^{d}}\frac{g^4_3}{2} \d^{ab} \int_0^\infty d t_1
\int_0^{\infty} d t_2 \int_0^{\infty} d t_3
\int_0^{t_2} d t_{z_1}
\int_0^{t_{z_1}} d t_{z_2} \, {\rm e}^{-m^2 (t_1 + t_2 + t_3)} \nl
& \times & \Delta^{-d/2} \exp{\left[p_1 \cdot p_2 \, 
G_c (t_1, t_2, t_3, t_{z_1}, t_{z_2})
\right] }~~,
\label{2ffin}
\eeqa
where
\beq
G_c (t_1, t_2, t_3, t_{z_1}, t_{z_2}) = \Delta^{-1} (t_{z_1} - t_{z_2}) 
\Big(\Delta - (t_1 + t_3) (t_{z_1} - t_{z_2}) \Big)~.
\label{gfish}
\eeq
This detailed analysis shows that, in the scalar case, the final form of 
the whole calculation can be simply determined by picking one of the various 
corners of integrations that are relevant for each diagram. All the other 
contributions only transform the regions of integration into the 
expected ones. This step can also be done by hand, by counting the number 
of configurations contributing to the diagram, and dividing it by the 
factor of $1/3!$ needed to perform the integration independently. 
However, if the symmetrization is carried out explicitly, it gives a 
non--trivial check on the correctness of the result, that can be useful
when computing more complicated amplitudes.

\begin{figure}[ht] 
\begin{center} 
{\scalebox{0.55}{\includegraphics{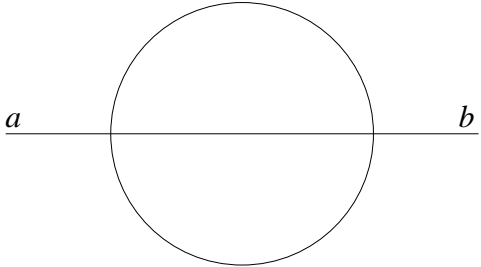}}}
\end{center}
 \caption{Two--loop two--point diagram in  the $\Phi^4$ theory.}
\label{sunset}
\end{figure}

We turn very briefly to the case of quartic interactions, by considering
the diagram in Fig.~\ref{sunset}.
The simplest way to get the correct result for this diagram is perhaps to use
the technique that was briefly mentioned in \secn{tree}, when the $\Phi^4$ 
matching was performed by inserting $\d$ functions for the proper times that 
must vanish in order to obtain quartic vertices. Choosing for example the
boundary $[A, A']$ in \fig{2lf}, one sees that the punctures must be 
integrated between $- 1/\sqrt{k_1}$ and $- \sqrt{k_1}$. After splitting 
the integration region into field theory propagators, as was done to obtain 
the $\Phi^3$ diagrams, we can simply insert the appropriate $\d$ functions,
say $\d((t_{z_1} - t_2)/\a')$, with strength $1/2$ since they are all located
at the boundaries of the integration regions. Four ways of inserting two 
such $\d$ functions yield an overall actor of unity. Including the other 
boundaries in the same way to complete the integration region yields 
\beq
A^{(2)}_2 = \frac{g^4}{(4 \pi )^d} N^2 \delta ^{a\, b}
	\int_0^{\infty} d t_1 \int_0^{\infty} d t_2 \int_0^{\infty} d t_3
	\Delta^{-d/2} e^{-m^2 (t_1 + t_2 + t_3) + 
	p^2 ~t_1 t_2 t_3/\Delta}~~,
\eeq
which is the leading color structure of the field theory diagram we wanted to
reproduce.

\subsection{Four--point amplitudes}

\begin{figure}[ht] 
\begin{center} 
{\scalebox{0.55}{\includegraphics{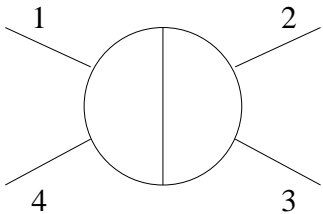}}}
\end{center}
 \caption{Two--loop four--point diagram in  the $\Phi^3$ theory.}
\label{twoloop4p}
\end{figure}

As a last non--trivial check of our technique for $\Phi ^3$ amplitudes,
we have computed the two--loop four--point diagram depicted in 
Fig.~\ref{twoloop4p}.
To extract this diagram from the general expression of the string master 
formula Eq.~(\ref{hmastac}), we can proceed in two steps:
first, since we are interested in an irreducible configuration,
we have to consider the limit $\eta _1 \to 0$ and introduce the change of
variables Eq.~(\ref{qvariable}), which generates the structure of the
irreducible two-loop vacuum bubble; then we have to consider all 
possible ways of inserting the four punctures. 
Two conditions must be respected: the cyclic order is fixed, say to 
$(1,2,3,4)$, and the pairs of external legs $(4,1)$ and $(2,3)$ must attach
to different propagators. For example, considering the boundary
corresponding to the interval $[A',A]$ in Fig.~\ref{2lf},
we can place 1 and 2  inside the interval $[-1,-\eta _1]$ and 2 and 3 outside,
or viceversa. Similar reasonings apply to the other two boundaries. Thus, 
introducing proper times for all the $z_i$ variables and summing up all 
the contributions, we obtain the expression
\beqa
A^{(2)}_4 (p_1, \ldots, p_4) & = & \frac{N^2}{2} 
{\rm Tr}\left( \lambda ^{a_1} \ldots \lambda ^{a_4} \right)
	\frac{g_{3}^6}{(4 \pi )^d}
	\int_0^{\infty }dt_1 \int_0^{\infty }dt_2 \int_0^{\infty }dt_3 
\label{fourpoint2} \\
& \times & \int_0^{t_3 }dt_{z_1} \int_0^{t_{z_1} }dt_{z_4}
	   \int_0^{t_1 }dt_{z_2} \int_0^{t_{z_2} }dt_{z_3}
	   {\rm e}^{- m^2(t_1 + t_2 + t_3)} \, \Delta ^{-d/2} \nl
& \times & \exp \left[ 
	p_1 \cdot p_2 G_a(t_2,t_1,t_3,t_{z_2},t_{z_1}) + 
	p_1 \cdot p_3 G_a(t_2,t_1,t_3,t_{z_3},t_{z_1}) \right. \nl
& &   + p_2 \cdot p_4 G_a(t_2,t_1,t_3,t_{z_2},t_{z_4}) +
	p_3 \cdot p_4 G_a(t_2,t_1,t_3,t_{z_3},t_{z_4}) \nl
& & \left. + p_2 \cdot p_3 G_c(t_2,t_1,t_3,t_{z_2},t_{z_3})
	   + p_1 \cdot p_4 G_c(t_2,t_3,t_1,t_{z_1},t_{z_4}) \right]~,
\nonumber
\eeqa
where the functions $G_a$ and $G_c$ have been defined in Eqs.~(\ref{gapple})
and~(\ref{gfish}) respectively. This result is exactly
the leading color term of the field theory diagram of Fig.~\ref{twoloop4p},
as expected.

\sect{Concluding remarks}
\label{concl}
\vskip 0.5cm

In this paper we analyze in detail the method to define and compute
field theory limits of string amplitudes and study in particular 
the case of scalar states. 
Conceptually, the target field theory is identified by isolating
in the string amplitude the quantities that
have to be kept fixed when $\a'$ goes to zero. This step is
essentially determined by looking at the simplest diagram, as was done
in \secn{tree}; in this way one can establish a  mapping between $g_S$
and the coupling constant of the field theory one wants to
reproduce. Having done this, it is possible to introduce
immediately the dimensional Schwinger parameters $t_i$ which measure
the length of the various propagators in units of $\a'$. This
operation absorbs the residual dependence on the string tension
contained in the overall normalization,  so that it becomes possible to
take the limit $\a' \to 0$, keeping the field coupling constant
and proper times $t_i$ fixed. The mapping between string logarithms
and Schwinger parameters is usually rather intuitive. As was stressed
through the various calculations, it is quite easy to follow, from a
geometrical point of view, how the string world--sheet degenerates
into a graph and how to relate each corner of the
integration region to a specific Feynman diagram.

In all our calculations, we have always found a precise matching
between the string results and the field theory Feynman diagrams. We
think that this work shows that string techniques are by now mature, and
can be confidently used to simplify the computation of many quantities
in field theory. The results obtained in this way can be firmly trusted; 
in fact, the derivation of the combinatorial factors is easier and the tedious
algebra of color decomposition is avoided. Moreover, during the
calculation itself, one has the possibility to perform novel 
consistency checks. For instance, in two--loop calculations one has 
to obtain the same result from very different regions of the string 
moduli space. We have shown that this is necessary in order to 
reconstruct the region of integration over the Schwinger proper 
times that is expected from field theory.

The generalization of this approach to the physically interesting case
of Yang--Mills theory is, of course, computationally more
demanding. One has to deal with derivatives of the Green
function, and all stringy quantities have to be expanded up to first
order in the multipliers, since the spin--1 particle is not the ground
state of the bosonic theory. Furthermore, besides this technical problem,
there is also a conceptual difference. As we discussed, the three
two--loop moduli $k_1$, $k_2$ and $\eta$ are  on the same
footing from the field theory point of view; in fact, they are all
associated to $\Phi^3$ propagators. Their role in the Schottky
parametrization is, however, quite different, since only the $k$'s are really
multipliers, while $\eta$ is a fixed point. This asymmetry is not
evident in the study of $\Phi^3$ diagrams and one obtains directly from the
string expression  the correct results, with the expected
{\em field theory} symmetry. For instance, \eq{2ffin} is invariant
under the exchange of $t_1$ and $t_3$. However, it is already clear
from the study of Yang--Mills vacuum bubbles~\cite{maru}, that the
different origin of the various parameters in the Schottky description 
plays a non--trivial role in this more complicated case. We think that the
study of the world--sheet geometry will provide other useful information
for the derivation of pure glue amplitudes from the string master formula,
and we hope that our analysis is a further step in this direction.

\vskip 1.5cm

{\large {\bf {Acknowledgements}}}
\vskip 0.5cm

We would like to thank R. Marotta and F. Pezzella for presenting
to us their results about $\Phi^4$ theory before publication, and for many
instructive discussions. We wish also to thank P. Di Vecchia, A. Lerda 
and S. Sciuto for many useful discussions and suggestions.
This work has been partially supported by the EEC under TMR contract
ERBFMRX-CT96-0045 and by the Fond National Suisse.

\vskip2cm

\appendix{\Large {\bf {Appendix}}}
\label{app}
\vskip 0.5cm
\renewcommand{\theequation}{A.\arabic{equation}}
\setcounter{equation}{0}

We collect in this Appendix the Feynman rules for the scalar theories
we are considering, our conventions and several useful formulas
concerning the computation of color factors.

The Feynman rule for the cubic scalar vertex described by the
lagrangian in \eq{lag3} is simply
\beq
V_{\a \b \c} = {\rm i} ~g_3 ~d_{\a \b \c}~~,
\label{v3}
\eeq
where $d_{\a \b \c}$ is the completely symmetric $U(N)$
color tensor, described below. Similarly, for the quartic interaction
in \eq{lag4}, the rule is
\beq
V_{\a \b \c \d} = {\rm i} ~g_4 \left(d_{\a \b \mu} d^\mu_{\; \, \c \d} +
d_{\a \c \mu} d^\mu_{\; \, \b \d} + d_{\a \d \mu} d^\mu_{\; \, \b \c} 
\right)~~.
\label{v4}
\eeq
Notice that in field theory we use the standard metric $(+,-,-,-)$,
whereas string theory is naturally formulated in the metric with
opposite sign.

To derive the $U(N)$ color algebra, it is useful to start from $SU(N)$
matrices, and then complement them with the diagonal $U(1)$
generator. In the following we will denote $U(N)$ indices with greek
letters, $\{ \a, \b, \ldots \}$, and $SU(N)$ indices with latin
ones, $\{a, b, \ldots \}$, so that, say, $\a = \{0,a\}$ if we
assign the value $0$ to the $U(1)$ index. Throughout the paper, most of the 
calculations have been performed with external particles restricted to 
the $SU(N)$ subgroup, unless explicitly stated. We normalize our generators
as
\beq
{\rm Tr} \left(\l_\a \l_\b \right) = \frac{1}{2} \d_{\a \b}~~.
\label{normgen}
\eeq
With this normalization, the $SU(N)$ generators satisfy
\beqa
\l_a \l_b & = & \frac{1}{2} \left[\frac{1}{N} \d_{a b} {\bf 1} +
\left(d_{a b c} + {\rm i} f_{a b c} \right) \l^c \right] \nl
\left( \l_a \right)^i_{\, j} \left( \l^a \right)^k_{\, l}
& = & \frac{1}{2} \left( \d^i_{\, l} \d^k_{\, j}  - 
\frac{1}{N} \d^i_{\, j} \d^k_{\, l} \right)~~,
\label{sun}
\eeqa
where $f_{a b c}$ are the $SU(N)$ structure constants, while $d_{a b
c}$ is the completely symmetric $SU(N)$ color tensor, satisfying
\beqa
d_{a c d} ~d_b^{\; \; c d} & = & \frac{N^2 - 4}{N} ~\d_{a b} \nl
d_{a e f} ~d_{\; \; b g}^f ~d_{\; \; \; \, c}^{g e} & = & 
\frac{N^2 - 12}{2 N} ~d_{a b c}~~.
\label{dsun}
\eeqa
To promote the above equations to $U(N)$ we must add a correctly
normalized $U(1)$ generator, which can be taken proportional to the
identity matrix,
\beq
\l_0 = \frac{1}{\sqrt{2 N}} {\bf 1}~~.
\label{l0}
\eeq
The anticommutation relations for the $U(N)$ generators can then be
summarized by
\beq
\left\{ \l_\a, \l_\b \right\} = d_{\a \b \c} \l^\c~~,
\label{acomm}
\eeq
provided one defines
\beqa
d_{a b 0} & = & \sqrt{\frac{2}{N}} \d_{a b} \nl
d_{a 0 0} & = & 0 \\
d_{0 0 0} & = & \sqrt{\frac{2}{N}}~~. \nonumber
\label{d0}
\eeqa
Note that this implies $d_{\a \b 0} = \sqrt{\frac{2}{N}} \d_{\a
\b}$, as well as $d_{a \a}^{\; \; \; \a} = 0$, extending to $U(N)$ the
corresponding $SU(N)$ property.
Using \eq{d0}, as well as the fact that $f_{a b 0} = 0$, one can
generalize Eqs.~(\ref{sun}) and (\ref{dsun}) to
\beqa
\l_\a \l_\b & = & \frac{1}{2} \left(d_{\a \b \c} + {\rm i} f_{\a \b
\c} \right) \l^\c \nl
\left( \l_\a \right)^i_{\, j} \left( \l^\a \right)^k_{\, l}
& = & \frac{1}{2} \left( \d^i_{\, l} \d^k_{\, j} \right)~~,
\label{un}
\eeqa
and
\beqa
d_{a \c \d} ~d_b^{\; \; \c \d} & = & N \, \d_{a b} \nl
d_{a \b \c} ~d_{\; \; b \d}^\c ~d_{\; \; \; \, c}^{\d \b} & = & 
\frac{N}{2} \, d_{a b c}~~.
\label{dun}
\eeqa
It is worth noticing however that \eq{dun} does not smoothly
generalize to the case in which one or more of the external indices
take their values in the $U(1)$ subgroup. For example one finds
\beqa
d_{0 \c \d} ~d_0^{\; \; \c \d} & = & 2 N \nl
d_{\a \b \c} ~d^{\a \b \c} & = & N (N^2 + 1) \\
d_{0 \a \b} ~d_{\; \; b \c}^\b ~d_{\; \; \; \, c}^{\c \a} & = & 
N \, d_{0 b c} \nl
d_{0 \a \b} ~d_{\; \; 0 \c}^\b ~d_{\; \; \; \, 0}^{\c \a} & = & 
2 N \, d_{0 0 0}~~. \nonumber
\label{dun0}
\eeqa
Finally, a useful formula to connect between standard Feynman rules
and color ordered ones is
\beq
d_{\a \b \c} = 2 ~{\rm Tr} \left( \l_\a \left\{ \l_\b, \l_\c \right\}
\right)~~.
\label{dtr}
\eeq
The analogous formula for the structure constants $f_{a b c}$ serves
the same purpose in QCD.

\vskip 2cm


\begin{thebibliography}{99}

\bibitem{scherk71} 
J. Scherk, {\it Nucl. Phys.} {\bf B 31} (1971) p. 222.

\bibitem{nesche72} 
A. Neveu and J. Scherk, {\it Nucl. Phys.} {\bf B 36} (1972) p. 55.

\bibitem{yo73} 
T. Yoneya, {\it Nuov. Cim. Lett.} {\bf 8} (1973) p. 951.

\bibitem{scheschw74} 
J. Scherk, J.H. Schwarz, {\it Nucl. Phys.} {\bf B 81} (1974) p. 118.

\bibitem{mets} 
R.R. Metsaev and A.A. Tseytlin, {\it Nucl. Phys.} {\bf B 298} (1988) p. 109.

\bibitem{kap88} 
V.S. Kaplunovsky, {\it Nucl. Phys.} {\bf B 307} (1988) p. 145, 
{\tt hep-th/9205068}.

\bibitem{kap92} 
V.S. Kaplunovsky, {\it Nucl. Phys.} {\bf B 382} (1992) p. 436, 
{\tt hep-th/9205070}.

\bibitem{beko91}
Z. Bern and D.A. Kosower,
{\it Phys. Rev. Lett.} {\bf 66} (1991) 1669.

\bibitem{beko92} 
Z. Bern and D.A. Kosower, {\it Nucl. Phys.} {\bf B 379} (1992) p. 451. 

\bibitem{be92} 
Z. Bern, {\it Phys. Lett.} {\bf B 296} (1992) p. 85.

\bibitem{bediko93}
Z. Bern, L. Dixon and D.A. Kosower,
{\it Phys. Rev. Lett.} {\bf 70} (1993) p. 2677, {\tt hep-ph/9302280}.

\bibitem{bernrev}
Z. Bern, L. Dixon and D.A. Kosower, {\it Ann. Rev. Nucl. Part. Sci.} 
{\bf 46} (1996) p. 109 {\tt hep-ph/9602280}.

\bibitem{beko88}
Z. Bern and D.A. Kosower, {\it Phys. Rev.} {\bf D 38} (1988) 1888.

\bibitem{letter} 
P. Di Vecchia, A. Lerda, L. Magnea and R. Marotta, {\it Phys. Lett.} 
{\bf 351 B} (1995) p. 445, {\tt hep-th/9502156}. 

\bibitem{big} 
P. Di Vecchia, L. Magnea, A. Lerda, R. Russo and R. Marotta,
{\it Nucl. Phys.} {\bf B469} (1996) p. 235, {\tt hep-th/9601143}.

\bibitem{bedushi}
Z. Bern, D.C. Dunbar and T. Shimada,
{\it Phys. Lett.} {\bf B 312} (1993) p. 277, {\tt hep-th/9307001}.

\bibitem{bddpr}
Z. Bern, L. Dixon, D.C. Dunbar, M. Perelstein and J.S. Rozowsky,
{\it Nucl. Phys.} {\bf B 530} (1998) p. 401, {\tt hep-th/9802162}.

\bibitem{kaj} 
K. Roland, {\it Phys. Lett.} {\bf 289 B} (1992) p. 148.

\bibitem{2loop}
P. Di Vecchia, L. Magnea, A. Lerda, R. Marotta and R. Russo,
{\it Phys. Lett.} {\bf B388} (1996) p. 65, {\tt hep-th/9607141}.

\bibitem{rolsat1}
K. Roland and H. Sato, {\it Nucl. Phys.} {\bf B 480} (1996) p. 99,
{\tt hep-th/9604152}.

\bibitem{rolsat2}
K. Roland and H. Sato, {\it Nucl. Phys.} {\bf B 515} (1998) p. 488,
{\tt hep-th/9709019}.

\bibitem{maru} L. Magnea and R. Russo, in {it Proceedings} of ``DIS 97'',
Chicago, USA, 1997, eds. J. Repond and D. Krakauer, AIP Conf. Proc. n. 407, 
p. 913, {\tt hep-ph/9706396}; also in {it Proceedings} of ``Beyond the 
Standard Model V'', Balholm, Norway, 1997, eds. G. Eigen, P. Osland and 
B. Stugu, AIP Conf. Proc. n. 415, p. 347, {\tt hep-ph/9708471}.

\bibitem{berntwol}
Z. Bern, J.S. Rozowsky and B. Yan, {\it Phys. Lett.} {\bf B 401} (1997)
p. 273, {\tt hep-ph/9702424}; also in {it Proceedings} of ``DIS 97'',
Chicago, USA, 1997, eds. J. Repond and D. Krakauer, AIP Conf. Proc. n. 407,
p. 908, {\tt hep-ph/9706392}. 

\bibitem{rolpa}
A. Pasquinucci, K. Roland, {\it Nucl. Phys.} {\bf B 485} (1997) p. 241,
{\tt hep-th/9608022}.

\bibitem{nap}
L. Cappiello, R. Marotta, R. Pettorino and F. Pezzella, {\it Mod. Phys. Lett.}
{\bf A13} (1998) p. 2433, {\tt hep-th/9804032}; 
L. Cappiello, R. Marotta, R. Pettorino and F. Pezzella, {\it Mod. Phys. Lett.}
{\bf A13} (1998) p. 2845, {\tt hep-th/9808164};
A. Liccardo, R. Marotta and F. Pezzella, {\it Mod. Phys. Lett.}
{\bf A14} (1999) p. 799, {\tt hep-th/9903027}.  

\bibitem{bedu} 
Z. Bern and D.C. Dunbar, {\it Nucl. Phys.} {\bf B 379} (1992) p. 562.

\bibitem{strass} 
M. J. Strassler, {\it Nucl. Phys.} {\bf B 385} (1992) p. 145, 
{\tt hep-ph/9205205}.

\bibitem{schu} 
M.G. Schmidt, C. Schubert, {\it Phys. Lett.} {\bf B 331} (1994)
p. 69, {\tt hep-th/9403158}.

\bibitem{schu2}
C. Schubert,
{\it Acta Phys. Polon.} {\bf B 27} (1996) p. 3965, {\tt hep-th/9610108}.

\bibitem{schsat1}
H. Sato and M.G. Schmidt, {\it Nucl. Phys.} {\bf B 524} (1998) p. 742,
{\tt hep-th/9802127}.

\bibitem{schsat2}
H. Sato and M.G. Schmidt, {\it Nucl. Phys.} {\bf B 560} (1999) p. 551,
{\tt hep-th/9812229}.

\bibitem{dfls88}
P. Di Vecchia, M. Frau, A. Lerda and S. Sciuto, {\it Nucl. Phys} {\bf B 298}
(1988) p.526.

\bibitem{d92} 
See, for example, P. Di Vecchia, {\it ``Multiloop amplitudes in string
theory''} in Erice, {\it Theor. Phys.} (1992), p.16, and references therein. 

\bibitem{MP}
R. Marotta and F. Pezzella, {\tt hep-th/9912158}.

\bibitem{scho} 
P. Di Vecchia, F. Pezzella, M. Frau, K. Hornfeck, A. Lerda and
S. Sciuto, {\it Nucl. Phys.} {\bf B 322} (1989) p. 317. 

\bibitem{GSW} 
M.B. Green, J.H. Schwarz and E. Witten, 
{\it ``Superstring Theory''}, Cambridge University Press (1987).

\bibitem{aalo}
V. Alessandrini, D. Amati, M. Le Bellac and D. Olive, {\it Phys. Rep.} {\bf 1}
(1971) p. 269.

\bibitem{kajun} 
K.~Roland, 
\newblock {\em Multiloop Amplitudes in Pure Gauge Theories: The
superstring Approach}
{SISSA/ISAS 131-93-EP} (1993).


\end{thebibliography}

\end{document}